\documentclass[12pt]{article}
\usepackage{amsthm,amsmath,latexsym,amssymb,amsfonts,amscd}
\usepackage{graphics,lscape,fancyhdr,array,stmaryrd,euscript,wrapfig}
\usepackage{empheq,slashed}
\usepackage{verbatim,slashed}
\numberwithin{equation}{section}
\usepackage{hyperref,setspace}
\usepackage{tikz-cd}
\usepackage{mathrsfs}
\usepackage[numbers,sort&compress]{natbib}
\usepackage[nottoc]{tocbibind}

\newcommand{\pl}{\partial}

\newcommand{\be}{\begin{align}}
\newcommand{\ee}{\end{align}}

\newcommand{\aA}{{\ensuremath{\mathcal{A}}}}
\newcommand{\aB}{{\ensuremath{\mathcal{B}}}}
\newcommand{\aC}{{\ensuremath{\mathcal{C}}}}
\newcommand{\aD}{{\ensuremath{\mathcal{D}}}}

\newcommand{\bry}{{{\bar{y}}}}
\newcommand{\brz}{{{\bar{z}}}}

\newcommand{\hs}{{\mathfrak{hs}}}

\newcommand{\fud}[2]{{}^{#1}{}_{#2}\,}
\newcommand{\fdu}[2]{{}_{#1}{}^{#2}\,}
\newcommand{\fudu}[3]{{}^{#1}{}_{#2}{}^{#3}\,}

\newcommand{\besubeqs}{\begin{subequations}}
\newcommand{\esubeqs}{\end{subequations}}

\usepackage{tensor}
\usepackage{todonotes}

\renewcommand{\bar}[1]{\overline{#1}}

\newtheorem{theorem}{\fbox{\color{violet}{Theorem}}}[section]

\newcommand{\TikzRect}[2]{\filldraw[color=black,fill=red]  (#1-\R,#2-\R) rectangle (#1+\R,#2+\R);}

\newcommand{\TikzRectG}[2]{\filldraw[color=black,fill=green]  (#1-\R,#2-\R) rectangle (#1+\R,#2+\R);}

\newcommand{\FIELDS}{
\begin{tikzpicture}[scale=0.7]
\tikzset{%
  >=latex, 
  inner sep=0pt,%
  outer sep=2pt,%
  mark coordinate/.style={inner sep=0pt,outer sep=0pt,minimum size=3pt,
    fill=black,circle}%
}
\def\R{0.15}
\def\mar{0.15}

\draw[->,black,thick] (0,0) -- (0,9) node[left]{$\# A$ };
\draw[->,black,thick] (0,0) -- (11,0) node[right]{$\# A'$} ;

\filldraw[color=black,fill=green]  (12,6) circle (\R);
\node[right] at (12.3,6) {one-forms, $\omega$};

\TikzRectG{12}{7}  \TikzRect{12.5}{7}  
\node[right] at (12.7,7) {zero-forms, $C$};

\filldraw[color=black,fill=green]  (4,0) circle (\R);
\filldraw[color=black,fill=green]  (3,1) circle (\R);
\filldraw[color=black,fill=green]  (2,2) circle (\R);
\filldraw[color=black,fill=green]  (1,3) circle (\R);
\filldraw[color=black,fill=green]  (0,4) circle (\R);
\draw[green,thick] (4,0) -- (0,4);

    \draw[-latex,thick] (4.5,5.5) node[right,scale=1.0]{$\omega^{A(2s-2)}$}
        to[out=-180,in=50] (+\mar, 4);
        
    \draw[-latex,thick] (4.5,8) node[right,scale=1.0]{$\Psi^{A(2s)}$, $\lambda=-s$ Weyl tensor}
        to[out=-120,in=10] (+\mar,5);

    \draw[-latex,thick] (8.5,1.5) node[right,scale=1.0]{$C^{A'(2s)}$, $\lambda=+s$ Weyl tensor}
        to[out=180,in=-45] (5, -\mar);
        
    \draw[rounded corners] ( -0.5 , 3.5 ) rectangle (0.5, 5.5) {};        

    \draw[-latex,thick] (2,-1) node[left,scale=1.0]{$\omega^{A'(2s-2)}$}
        to[out=0,in=-90] (4, -\mar);

    \TikzRect{0}{5}
    \TikzRect{1}{6}
    \TikzRect{2}{7}
    \TikzRect{3}{8}
    \draw[red,thick] (0,5) -- (4,9);

    \TikzRectG{5}{0}
    \TikzRectG{6}{1}
    \TikzRectG{7}{2}
    \TikzRectG{8}{3}
    \draw[green,thick] (5,0) -- (9,4);

    \filldraw[color=black,fill=black]  (4.5-1.5*\R,-\R) rectangle (4.5+1.5*\R,+\R);

    \draw[-latex,thick] (4.5,3.5) node[above,scale=1.0]{\large $\mathcal{V}(e,e,C)$-cocycle}
        to[out=-90,in=90] (4.5, 0.5-\mar);

\end{tikzpicture}}

\newtheorem{lemma}[theorem]{Lemma}

\newtheorem{definition}[theorem]{Definition}

\usepackage[all,cmtip]{xy}

\begin{document}
\pagenumbering{gobble}
\hfill
\vskip 0.01\textheight
\begin{center}
{\Large\bfseries 
Minimal Model of Chiral Higher Spin Gravity}

\vspace{0.4cm}

\vskip 0.03\textheight
\renewcommand{\thefootnote}{\fnsymbol{footnote}}
Alexey \textsc{Sharapov}${}^{a}$, 
Evgeny \textsc{Skvortsov}\footnote{Research Associate of the Fund for Scientific Research -- FNRS, Belgium}${}^{b,c}$, Arseny \textsc{Sukhanov}${}^d$ \& Richard \textsc {Van Dongen}${}^{b}$
\renewcommand{\thefootnote}{\arabic{footnote}}
\vskip 0.03\textheight

{\em ${}^{a}$Physics Faculty, Tomsk State University, \\Lenin ave. 36, Tomsk 634050, Russia}\\
\vspace*{5pt}
{\em ${}^{b}$ Service de Physique de l'Univers, Champs et Gravitation, \\ Universit\'e de Mons, 20 place du Parc, 7000 Mons, 
Belgium}\\
\vspace*{5pt}
{\em ${}^{c}$ Lebedev Institute of Physics, \\
Leninsky ave. 53, 119991 Moscow, Russia}\\
\vspace*{5pt}
{\em ${}^{d}$
Moscow Institute of Physics and Technology, \\
Institutskiy per. 7, Dolgoprudnyi, 141700 Moscow region, Russia}

\end{center}

\vskip 0.02\textheight

\begin{abstract}
A unique class of local Higher Spin Gravities with propagating massless fields in $4d$ --- Chiral Higher Spin Gravity --- was first found in the light-cone gauge. We construct a covariant form of the corresponding field equations in all orders, thus completing the previous analysis of \href{https://arxiv.org/abs/2204.10285}{arxiv:2204.10285}. This result is equivalent to taking the minimal model (in the sense of $L_\infty$-algebras) of the jet-space BV-BRST formulation of Chiral Higher Spin Gravity, thereby, containing also information about counterterms, anomalies, etc.
\end{abstract}

\newpage
\tableofcontents
\newpage
\section{Introduction}
\label{sec:}
\pagenumbering{arabic}
\setcounter{page}{2}
Higher Spin Gravities (HiSGRA's) --  theories that extend gravity with massless higher spin fields \cite{Bekaert:2022poo} -- are difficult to construct, which, as can be argued, is due to the fact that the masslessness shifts some of the genuine quantum gravity UV problems down to the classical level.  Therefore, it should not come as a surprise that there is a handful of HiSGRA's available at the moment: $3d$ topological theories with (partially-)massless and conformal fields \cite{Blencowe:1988gj,Bergshoeff:1989ns,Campoleoni:2010zq,Henneaux:2010xg,Pope:1989vj,Fradkin:1989xt,Grigoriev:2019xmp}; $4d$ conformal HiSGRA \cite{Segal:2002gd,Tseytlin:2002gz,Bekaert:2010ky} that is a higher spin extension of Weyl gravity; Chiral HiSGRA \cite{Metsaev:1991mt,Metsaev:1991nb,Ponomarev:2016lrm,Skvortsov:2018jea,Skvortsov:2020wtf} and its contractions \cite{Ponomarev:2017nrr,Krasnov:2021nsq}. Nevertheless, the infinite-dimensional gauge symmetries associated with higher spin fields are believed to be sufficient to eliminate all counterterms.

Chiral HiSGRA is a unique Lorentz invariant and local completion that contains at least one massless higher spin field with a nontrivial self-interaction. The theory was first found in the light-cone gauge. It is most easily described with the help of its cubic amplitudes thanks to the close relation between the light-cone gauge and spinor-helicity formalism \cite{Chalmers:1998jb,Chakrabarti:2005ny,Chakrabarti:2006mb,Bengtsson:2016jfk,Ponomarev:2016cwi}. Indeed, for any triplet of helicities such that $\lambda_1+\lambda_2+\lambda_3>0$ there is a unique vertex and the corresponding amplitude \cite{Bengtsson:1986kh,Benincasa:2011pg}: 
\begin{align}\label{genericV}
   V_{\lambda_1,\lambda_2,\lambda_3}\Big|_{\text{on-shell}} \sim 
    [12]^{\lambda_1+\lambda_2-\lambda_3}[23]^{\lambda_2+\lambda_3-\lambda_1}[13]^{\lambda_1+\lambda_3-\lambda_2}\,.
\end{align}
The minimal assumptions stated above forces one \cite{Metsaev:1991mt,Metsaev:1991nb,Ponomarev:2016lrm} to introduce fields of all spins, including the scalar and spin-two, i.e. graviton, and fix the couplings constants to be 
\begin{align}\label{eq:magicalcoupling}
    V_{\text{Chiral}}&= \sum_{\lambda_1,\lambda_2,\lambda_3}  C_{\lambda_1,\lambda_2,\lambda_3}V_{\lambda_1,\lambda_2,\lambda_3}\,, && C_{\lambda_1,\lambda_2,\lambda_3}=\frac{\kappa\,(l_p)^{\lambda_1+\lambda_2+\lambda_3-1}}{\Gamma(\lambda_1+\lambda_2+\lambda_3)}\,.
\end{align}
Here $l_p$ is a constant of dimension length and $\kappa$ is an arbitrary dimensionless constant. 

In the present paper, we complete the program started in \cite{Skvortsov:2022syz} -- to construct manifestly Lorentz covariant equations of motion of Chiral Theory in the form of a Free Differential Algebra (FDA). This can also be understood as the minimal model of the jet space BV-BRST complex of Chiral HiSGRA. Therefore, and this is the main motivation to look for this FDA, the result contains the same information as the local BRST cohomology, see e.g. \cite{Barnich:1994db,Barnich:1994mt}. 

We refer to \cite{Skvortsov:2022syz} for more details regarding the context and notation. The outline of the present paper is as follows. In Section \ref{sec:initial}, we review briefly the minimal models and FDA's \cite{Sullivan77, vanNieuwenhuizen:1982zf,DAuria:1980cmy} together with specifics of HiSGRA's applications \cite{Vasiliev:1988sa}. In Section \ref{sec:free}, we recall how to formulate free Chiral HiSGRA as an FDA. The main result is in Section \ref{sec:FDA}, where, after recalling the first few vertices found in \cite{Skvortsov:2022syz}, we propose an appropriate completion to all orders.

\section{Minimal models}
\label{sec:initial}
In this paper, we are seeking for a covariant form of Chiral Theory at the level of equations of motion. The first few orders were constructed in \cite{Skvortsov:2022syz}. Looking for a quantum gravity model in the form of classical equations of motion may seem like an over-simplification since the equations do not help much with quantization. While in general this is true, it is worth noting that there exist methods to quantize non-Lagrangian systems \cite{Kazinski:2005eb} and they require some additional structures on top of equations of motion, see also \cite{Misuna:2020fck,Sharapov:2021drr}. However, Chiral Theory does have a simple action in the light-cone gauge. For a given triplet of helicities $\lambda_{1,2,3}$ the cubic vertex involves $\lambda_1+\lambda_2+\lambda_2>0$ derivatives. The challenge is to find a covariant form of this theory. Interactions with one and two derivatives are of Yang--Mills and gravitational type and can be covariantized without much trouble \cite{Krasnov:2021nsq}. In a sense, certain first order actions allow one to represent such interactions without having to introduce (many) auxiliary fields to encode higher derivatives. For all other interactions, i.e. those with more than two derivatives, a clever choice of auxiliary fields is needed. Auxiliary fields should be introduced consistently with (infinite-dimensional higher spin) gauge symmetries, which is hard. 

Equations of motion are simpler, of course, but they lack a lot of important information. We utilize an approach which is not fraught with such pitfalls. The idea is to look for a (non-negatively graded) supermanifold $\mathcal{N}$ equipped with an odd nilpotent vector field $Q$:
$$
|Q|=1\,,\qquad [Q,Q]=2Q^2=0\,.
$$
Such a vector field is called usually homological and the pair $(\mathcal{N},Q)$ is referred to as a differential graded manifold.   Given these data, one can write down a sigma-model:  
\begin{align}\label{mevenmore}
    d \Phi&= Q(\Phi) \,,
\end{align}
where the form-fields $\Phi = \Phi(x,dx)$ define and are defined by the maps  
$\Phi: T[1]\mathcal{M} \rightarrow \mathcal{N}$ from the odd tangent bundle of a space-time manifold $\mathcal{M}$ to the target space $\mathcal{N}$. As a historical comment, equations of the form \eqref{mevenmore} were introduced under the name Free Differential Algebras (FDA) by Sullivan \cite{Sullivan77} to study rational homotopy theory. The notion of the minimal model also appeared in \cite{Sullivan77} in the case of differential graded algebras. Since differential forms of various degrees are naturally incorporated into FDA, it was used in the context of supergravities \cite{vanNieuwenhuizen:1982zf,DAuria:1980cmy}. Later, FDA was applied to the higher spin problem \cite{Vasiliev:1988sa}, see also \cite{Vasiliev:2005zu,Barnich:2010sw} and \cite{Barnich:2005ru} for the relation between AKSZ \cite{Alexandrov:1995kv} and FDA.

The question is how to find $\mathcal{N}$ and $Q$ such that \eqref{mevenmore} together with natural gauge symmetries are equivalent to the sought for equations of motion. Since the coordinates on $\mathcal{N}$ are determined by the free limit, i.e. are indifferent to interactions, we will have exactly the same fields/coordinates as \cite{Vasiliev:1988sa}. Another question is how much information about the theory is stored in $Q$. It turns out \cite{Barnich:2010sw,Grigoriev:2012xg,Grigoriev:2019ojp,Grigoriev:2020lzu} that such $Q$ is closely related to the jet space BV-BRST formulation of the original field theory (whenever it is already available) \cite{Brandt:1997iu,Brandt:1996mh, Kaparulin:2011xy,Barnich:1994db,Barnich:1994mt}. To be a bit more precise, the jet space BV-BRST complex gives a (usually very large) $L_\infty$-algebra. One can consider various (quasi-isomorphic) reductions of this $L_\infty$-algebra. The smallest one is called the minimal model. Some care is needed to extend this notion to field theories \cite{Barnich:2009jy,Grigoriev:2019ojp}. The minimal model contains exactly the same information (locally) as the original BV-BRST formulation \cite{Barnich:2009jy}, e.g. actions (in the sense of the on-shell symplectic structure), counterterms, anomalies, charges etc. correspond to various $Q$-cohomology. 

Therefore, Eq. \eqref{mevenmore} is more useful than it seems, encoding  the same information as local BRST cohomology. As the BV-BRST formulation of (covariantized) Chiral Theory is not yet available, a pragmatic point of view is to directly look for the minimal model and $Q$. The coordinates on $\mathcal{N}$ are easy to find by looking at the free field limit. Some essential information about covariant interactions was found in \cite{Krasnov:2021nsq,Skvortsov:2022unu,Skvortsov:2022syz}. For majority of field theories, the coordinates on $\mathcal{N}$ divided into two groups: those of degree-one, which we denoted $\omega$, and those of degree-zero, denoted $C$. By abuse of notation we use the same symbols both for the coordinates on $\mathcal{N}$ and for the corresponding sigma-model fields. 

The simplest system with such initial data consists of the flatness condition for a connection $\omega$ of some Lie algebra together with the covariant constancy condition on a zero-form $C$ in some representation $\rho$:
\begin{align}
    d\omega &=\tfrac12[\omega,\omega]\,,& dC&=\rho(\omega)C \label{laxd}\,.
\end{align}
In the most general case we should have 
\begin{equation}\label{mostgeneral}
    \begin{array}{rcl}
         d\omega&=&l_2(\omega,\omega)+l_3(\omega,\omega,C)+l_4(\omega,\omega,C,C)+\ldots\,,\\
    dC&=&l_2(\omega,C)+l_3(\omega,C,C)+\ldots\,.
    \end{array}
\end{equation}
A useful interpretation of the nilpotency condition for $Q$
\begin{align}
    Q^2&=0 &&\Longleftrightarrow && Q^\aB \frac{\pl}{\pl \Phi^\aB} Q^\aA=0
\end{align}
is that it is equivalent to the relations that define an $L_\infty$-algebra \cite{Alexandrov:1995kv}. To derive the latter we represent $Q$ as
\begin{align}
    Q&= \sum_{n\geq2} l_n(\omega,\omega,C,...,C) \frac{\delta }{\delta \omega}+\sum_{n\geq2} l_n(\omega,C,...,C) \frac{\delta }{\delta C}\,.
\end{align}
In what follows, the $L_\infty$-algebra will originate from an $A_\infty$-one. Therefore, let us give a definition of $A_\infty$. $A_\infty$ is a graded vector space together with degree $(-1)$ maps $m_n(\bullet,\ldots,\bullet)$ that satisfy $m\circ m=0$. Here, $m=m_1+m_2+...$ is a formal sum and $\circ$ is the Gerstenhaber product, which is defined for any two multilinear maps $f$ and $g$ of degrees $|f|$ and $|g|$ and having $k_f$ and $k_g$ arguments as
\begin{align}\label{gersproduct}
    f\circ g&= \sum_i (-1)^\kappa f(a_1,\ldots,a_i, g(a_{i+1},\ldots,a_{i+k_g}),a_{i+k_g+1},\ldots, a_{k_f+k_g-1})   \,.
\end{align}
Here $\kappa$ is the usual Koszul sign: $\kappa=|g|(|a_1|+\cdots+|a_i|)$. Any $A_\infty$-algebra gives rise to an $L_\infty$-algebra via symmetrization of arguments.

Chiral Theory's FDA originates from an $A_\infty$-algebra that, as a graded space, consists of degree-one and degree-zero subspaces, $A=A_{0}\oplus A_{1}$. In practice, the difference between $L_\infty$ and $A_\infty$ structure maps is in that the former are graded symmetric while the latter do not have any symmetry properties and the order of arguments does matter, e.g. $m_2(a,u)$ and $m_2(u,a)$, $a\in A_{1}$, $u\in A_0$ are different maps. Once $A_\infty$ is available, we can construct the associated $L_\infty$. For example,  $m_2(a,u)$ and $m_2(u,a)$ give rise to $l_2(a,u)=m_2(a,u)+m_2(u,a)$, and so on. Therefore, the general strategy and techniques we use can be borrowed from \cite{Sharapov:2018ioy,Sharapov:2018kjz,Sharapov:2019vyd,Sharapov:2020quq,Sharapov:2022fos}.

\section{Free FDA}
\label{sec:free}
A useful starting point is to write down democratic equations proposed by Roger Penrose \cite{Penrose:1965am} to describe fields of definite helicity\footnote{We use the two-component indices $A,B,...=1,2$, $A',B',...=1,2$ of the Lorentz algebra $sl(2,\mathbb{C})$. The invariant tensors are $\epsilon^{AB}=-\epsilon^{BA}$ and $\epsilon^{A'B'}=-\epsilon^{B'A'}$, see \cite{penroserindler} and \cite{Skvortsov:2022syz} for more detail. }
\begin{align}\label{hsA}
    \nabla\fud{A}{B'} \Psi^{B'A'(2s-1)}&=0\,,
    &\nabla\fdu{B}{A'} \Psi^{BA(2s-1)}&=0\,.
\end{align}
Say, the first one describes helicity $(+s)$ and the second one helicity $(-s)$ states. In the case of $s=1$ one finds the Maxwell equations for the (anti-)self-dual components of $F_{\mu\nu}$. For $s=2$ the equations are Bianchi identities for the (anti-)self-dual components of the Weyl tensor $C_{\mu\nu,\lambda\rho}$. For arbitrary $s$ we will refer to the fields of \eqref{hsA} as Weyl tensors. Many interesting covariant interactions require gauge fields, e.g. $A_\mu$ and $g_{\mu\nu}$ for $s=1$ and $s=2$ in place of the Weyl tensors. The twistor approach is not an exception, but it requires to introduce a gauge potential only for one of the helicities, that is, the positive and negative helicities are treated differently \cite{Hughston:1979tq,Eastwood:1981jy,Woodhouse:1985id}. The gauge potential for the positive helicity field obeys
\begin{align}\label{hsB}
    \nabla\fud{A}{A'}\Phi^{A(2s-1),A'}&=0\,, && \delta\Phi^{A(2s-1),A'}=\nabla^{AA'}\xi^{A(2s-2)}\,,
\end{align}
and the corresponding Weyl tensor is $2s-1$ derivatives removed from the potential: $\Psi^{A'(2s)}\sim \nabla\fdu{A}{A'}...\nabla\fdu{A}{A'}\Phi^{A(2s-1),A'}$. A more geometric description \cite{Hitchin:1980hp} is to take a one-form $\omega^{A(2s-2)}$ (to become a gauge connection later). It has one component more as compared to $\Phi^{A(2s-1),A'}$:
\begin{align}
\omega^{A(2s-2)}\equiv e_{BB'}\Phi^{A(2s-2)B,B'}+e\fud{A}{B'}\Theta^{A(2s-3),B'}\,,
\end{align}
where $e^{AA'}\equiv e^{AA'}_\mu\, dx^\mu$ is the vierbein one-form. Similarly to the vierbein, $\omega$ has an extra component and an extra algebraic symmetry is called for to ensure it does not contribute to dynamics. Altogether, the gauge symmetries read  
\begin{align}\label{lin-gauge}
    \delta \omega^{A(2s-2)}&= \nabla \xi^{A(2s-2)} +e\fud{A}{C'} \eta^{A(2s-3),C'}\,,
\end{align}
There is a simple action \cite{Hitchin:1980hp} that leads to equations equivalent to \eqref{hsB} and to the second of \eqref{hsA}. In terms of the new variables it acquires a very compact form \cite{Krasnov:2021nsq}:
\begin{align}\label{niceaction}
    S= \int \Psi^{A(2s)}\wedge H_{AA}\wedge \nabla \omega_{A(2s-2)}\,.
\end{align}
Here $H^{AB}\equiv e\fud{A}{C'}\wedge e^{BC'}$. The action is consistent on self-dual backgrounds, i.e. it already goes beyond what is easy to achieve with the standard covariant approach where a spin-$s$ field is described \cite{Fronsdal:1978rb} by a rank-$s$ symmetric tensor $\Phi_{\mu_1...\mu_s}$. 

\begin{figure}[h!]
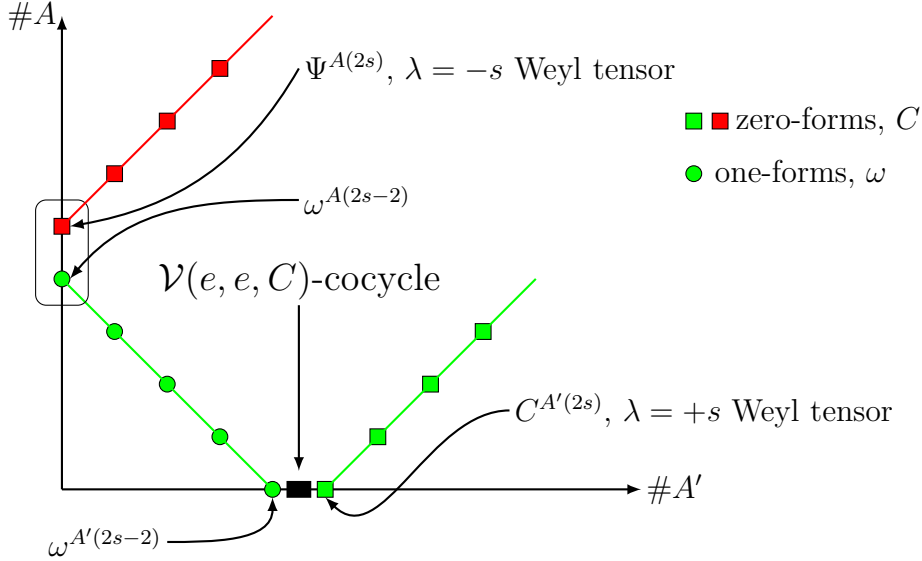

\FIELDS
  \caption{A diagram to show fields/coordinates involved into the description of higher spin fields. Along the axes we have the number of unprimed/primed indices on a spin-tensor. Degree-one coordinates are circles, degree-zero ones are rectangles. The fields that describe a helicity $\lambda=+s$ state are green. Those needed to describe helicity $\lambda=-s$ state are red. The black square shows a cocycle that links the one-form sector to zero-forms (at the free level it relates two fields for each spin's subsystem). The two fields in the rounded rectangle enter the free action. The rest of the fields encode derivatives thereof in a coordinate invariant and background independent way. The solid lines link pairwise the fields that `talk' to each other in the free equations. }\label{fig:figure1}
\end{figure}

\paragraph{Free equations of motion as Free Differential Algebras.} The variational equations do not have an FDA form yet:
\begin{align}\label{first}
    \nabla \Psi^{A(2s)}\wedge H_{AA}&=0\,, && H^{AA}\wedge \nabla \omega^{A(2s-2)}=0\,.
\end{align}
The l.h.s. represents a certain operator that constrains the form of $\nabla \Psi$ or $\nabla \omega$. The kernel can easily be parameterized by another field to recast the equation into an FDA form:
\begin{align}
    \nabla \Psi^{A(2s)}&= e_{BB'}\Psi^{A(2s)B,B'}\,,
&
    \nabla \omega^{A(2s-2)} &= e\fud{A}{B'} \omega^{A(2s-3),B'}\,,
\end{align}
where we introduced a zero-form $\Psi^{A(2s+1),A'}$ and a one-form $\omega^{A(2s-3),A'}$. This brings two new coordinates on $\mathcal{N}$ and  we need to extend $Q$ to those as well, and so on and so forth.\footnote{The entire field content is known since \cite{Vasiliev:1986td}. It is the equations that are different. } The final result of this process is ($\nabla^2=0$ in flat space)
\besubeqs\label{freeeq}
\begin{align}
    \nabla \omega^{A(n-i),A'(i)}&= e\fud{A}{B'} \omega^{A(n-i-1),A'(i)B'}\,, && i=0,...,n-1\,,\\
    \nabla \omega^{A'(n)}&= H_{B'B'} C^{A'(n)B'B'}\,,\\
    \nabla C^{A(k),A'(n+k+2)}&= e_{BB'} C^{A(k)B,A'(n+k+2)B'}\,, && k=0,1,2,...\,,\\
    \nabla \Psi^{A(n+k+2),A'(k)}&= e_{BB'} \Psi^{A(k+n+2)B,A'(k)B'}\,, && k=0,1,2,...\,,
\end{align}
\esubeqs
where $C$ and $\Psi$ are zero forms and $\omega$ are one-forms. Figure \ref{fig:figure1} illustrates the field content for a fixed spin $s$. 
The only dynamical fields are the ones that enter the action. The rest of the fields are expressed as derivatives of the dynamical ones. It may not seem very useful to introduce infinitely many fields to encode higher derivatives of the dynamical ones, especially when the fields are free, but it can be of great help later: interactions of Chiral Theory may have any number of derivatives (with helicities $\lambda_i$ on external legs fixed, the number of derivatives is $\lambda_1+\lambda_2+\lambda_3>0$, hence, finite).
With the help of generating functions the equations can be rewritten in a more concise form. To this end, we introduce
\begin{align}
    \omega(y,\bry)&= \sum_{n,m}\tfrac{1}{n!m!} \omega_{A(n),A'(m)}\, y^A...y^A\, \bry^{A'}...\bry^{A'}
\end{align}
and pack both $C^{A(n+k+2),A'(k)}$ and $\Psi^{A(k),A'(n+k+2)}$ into a single generating function $C(y,\bry)$. On top of that $C(y,\bry)$ contains $C^{A(k),A'(k)}$ to describe a scalar field, which is necessarily present in Chiral Theory. The free equations now read 
\begin{align}\label{linearizeddata}
    \nabla\omega &= e^{BB'}y_{B} \pl_{B'} \omega +H^{B'B'} \pl_{B'}\pl_{B'}C(y=0,\bry)\,,& 
    \nabla C&= e^{BB'}\pl_B \pl_{B'} C\,.
\end{align}
(Recall that $\nabla^2=0$ in Minkowski space.)
These equations serve as a departing point for constructing a non-linear theory. Indeed, viewing \eqref{linearizeddata} as linearization of some theory of type 
\eqref{mostgeneral}, we can write
\begin{align}\label{linearizeddataA}
    d\omega &= \mathcal{V}(\omega_0, \omega) +\mathcal{V}(\omega_0,\omega_0,C)\,,& 
    d C&= \mathcal{U}(\omega_0, C)\,,
\end{align}
where the structure maps have some arguments set to their background value $\omega_0$ in Minkowksi space.

\section{Chiral FDA}
\label{sec:FDA}
We begin by recalling the general form of the FDA for Chiral Theory, boundary conditions and the main results of \cite{Skvortsov:2022syz}, i.e. the next-to-leading order (NLO) vertices. The main result of the present paper is the construction of all the higher order vertices that is presented step by step in Section \ref{sec:higherorders}. This result requires the tools of homological perturbation theory that go beyond the technique applied in \cite{Skvortsov:2022syz}. 

\subsection{General form, initial data, and boundary conditions}
\label{sec:}
The coordinates on $\mathcal{N}$ or, what is the same,  the spectrum of fields are determined by the free limit. Therefore, we are seeking for  a theory with fields given by generating functions $\omega(y,\bry)$ and  $C(y,\bry)$. Chiral Theory may have Yang--Mills gaugings \cite{Skvortsov:2020wtf} of Chan--Paton type, i.e. one can define $U(N)$, $O(N)$ and $USp(N)$ gaugings. To take this into account we assume that $\omega$ and $C$ take their values in $\mathrm{Mat}_N$.\footnote{As was shown in \cite{Sharapov:2018kjz,Sharapov:2019vyd,Sharapov:2020quq}, having matrix factors leads to great simplifications, reducing a complicated Chevalley-Eilenberg cohomology problem to a much simpler Hochschild one. } With the field content at hand, the most general ansatz for Chiral Theory's FDA reads
\besubeqs\label{eq:chiraltheory}
\begin{align} 
    d\omega&= \mathcal{V}(\omega, \omega) +\mathcal{V}(\omega,\omega,C)+\mathcal{V}(\omega,\omega,C,C)+...\,,\\
    dC&= \mathcal{U}(\omega,C)+ \mathcal{U}(\omega,C,C)+... \,.
\end{align}
\esubeqs
We recall that the structure maps originate from a certain $A_\infty$-algebra and for that reason the order of arguments does matter. At each order we can, in principle, have vertices with various positions of $\omega$ and $C$.  
Free equations \eqref{linearizeddata}, when interpreted as linerization \eqref{linearizeddataA} of \eqref{eq:chiraltheory} lead to a number of boundary conditions for the $L_\infty$ structure maps:
\besubeqs\label{eq:boundaryconditions}
\begin{align}
    \mathcal{V}(e,\omega)+\mathcal{V}(\omega,e)&=e^{CC'} y_C \pl_{C'} \omega\,,\\
    \mathcal{U}(e,C)+\mathcal{U}(C,e)&=e^{CC'} \pl_C \pl_{C'} C\,,\\
    \mathcal{V}(e,e,C)&= e\fdu{B}{C'}e^{BC'} \pl_{C'} \pl_{C'} C(y=0,\bry)\,.
\end{align}
\esubeqs
Here, $e=e^{AA'}\, y_A \bry_{A'}$. To summarize, we are looking for a theory with the spectrum of fields given in \eqref{freeeq}, in the form of FDA \eqref{eq:chiraltheory} such that it reproduces the boundary conditions \eqref{eq:boundaryconditions}, i.e. the free equations. 

\subsection{FDA up to NLO}
\label{sec:}
$L_\infty$ structure maps, known and to be found, can take a number of $\omega$- and $C$-type arguments. In terms of components they just contract a certain number of indices on each pair of arguments, leaving the other indices contracted with $y$ and $\bry$. This can be achieved by representing the structure maps by poly-differential operators:
\begin{align}
    \mathcal{V}(f_1,...,f_n)&= \mathcal{V}(y, \pl_1,...,\pl_2)\, f_1(y_1)...f_n(y_n) \Big|_{y_i=0}\,,
\end{align}
where $f_i$'s are $\omega$'s or $C$'s. We have explicitly indicated dependence of $y$'s; the $\bry$'s can be treated similarly. Our notation can be summarized as follows: (i) we abbreviate $y^{A}\equiv p_0^{A}$, $\pl^{y_i}_{A}\equiv p_{A}^i$, $\bry^{A'}\equiv r_0^{A'}$, $\pl^{\bry_i}_{A'}\equiv r_{A'}^i$; (ii) contractions $p_{ij}\equiv p_i \cdot p_j\equiv -\epsilon_{AB}p^A_{i}p_{j}^B=p^A_{i}p_{jA}$ are done in such a way that $\exp[p_0\cdot p_i]f(y_i)=f(y_i+y)$; (iii) Lorentz invariance forbids to mix primed and unprimed indices; (iv) all indices must be contracted with $\epsilon_{AB}$ or $\epsilon_{A'B'}$; (v) explicit arguments $y_i$ in $f$'s and the symbol  $|_{y_i=0}$ are omitted. Importantly, all operators are assumed to be local, i.e. they send polynomials to polynomials.\footnote{A crucial difference from \cite{Vasiliev:1988sa} is that there are additional restrictions we impose on the vertices to make them represent local and independent interactions, rather than infinite sums over equivalent interactions. The latter can lead to divergent observables \cite{Boulanger:2015ova,Skvortsov:2015lja}  and merely represent the most general ansatz for gauge-invariant interactions rather than a concrete field theory. }

After these preparations, we can rewrite the boundary conditions \eqref{eq:boundaryconditions} in the operator language as
\besubeqs\label{eq:boundaryconditionsB}
\begin{align}
    \mathcal{V}(e,\omega)+\mathcal{V}(\omega,e)&\sim p_{01}r_{12}\, e^{p_{02}+r_{02}}\, ( e^{CC'}y^1_C \bry^1_{C'})\, \omega(y_2,\bry_2)\Big|_{y_{1,2}=\bry_{1,2}=0}\label{eq:boundaryconditionsBA}\,,\\
    \mathcal{U}(e,C)+\mathcal{U}(C,e)&\sim r_{12} p_{12}\, e^{p_{02}+r_{02}}\, ( e^{CC'}y^1_C \bry^1_{C'})\, C(y_2,\bry_2)\Big|_{y_{1,2}=\bry_{1,2}=0}\label{eq:boundaryconditionsBB}\,,\\
    \mathcal{V}(e,e,C)&\sim  r_{13} r_{23} p_{12}\, e^{r_{03}}\,( e^{BB'}y^1_B \bry^1_{B'})( e^{CC'}y^2_C \bry^2_{C'})\, C(y_3,\bry_3)\Big|_{y_{1,2,3}=\bry_{1,2,3}=0}\label{eq:boundaryconditionsBC}\,,
\end{align}
\esubeqs
where  $\sim$ implies an unessential proportionality coefficient. The dependence of $\bry$'s will always be the same: all structure maps have factorized form
\begin{align}
    \mathcal{V}(f_1,...,f_n)&= v(f_1'(y),..., f_n'(y)) \otimes f_1''(\bry)\star... \star f_n''(\bry)  \,,
\end{align}
where $f_i(y,\bry)=f_i'(y) \otimes f_i''(\bry)$. 
All $\bry$-dependent factors are multiplied via the star-product:
\begin{align}
    f_1''(\bry)\star ...\star f_n''(\bry)&= \exp{\left[\sum_{0=i<j=n} r_i \cdot r_j\right]} f_1''(\bry_1)...f_n''(\bry_n)\Big|_{\bry_i=0}\,.
\end{align}
Therefore, we have the first Weyl algebra $A_1$ in $\bry$'s, which then combines nicely with (optional) matrix factors into a bigger associative algebra. It is the dependence of $y$'s that we will be concerned with. In what follows, we assume that any algebra or module always has a trivial, but very useful, factor of $A_1\otimes \mathrm{Mat}_N$, which, in fact, can be replaced by any associative algebra. In practice, this means that we cannot permute $\omega$ and $C$ factors and should treat the $L_\infty$-maps as $A_\infty$-maps. The latter implies that our $L_\infty$-algebra originates from an $A_\infty$-algebra via the symmetrization map.

\paragraph{Higher spin algebra.} The $A_\infty$-relations at order $\omega^3$ imply the associativity of $\mathcal{V}(\bullet,\bullet)$:
\begin{equation}\label{eq:omegacubed}
    \mathcal{V}(\mathcal{V}(a,b),c)-\mathcal{V}(a,\mathcal{V}(b,c))=0\,.
\end{equation}
The algebra is easy to find: its light-cone gauge version appeared in \cite{Ponomarev:2017nrr} and its covariant version was guessed in \cite{Krasnov:2021cva}. The algebra is just the commutative algebra $\mathbb{C}[y]$ of polynomial functions in $y$.\footnote{We recall temporarily that we will always forget the $A_1$ factor of $\bry$ and the matrix factors. } The symbol of the product is\footnote{We chose this product to have the simplest prefactor $1$ as compared to $1/2$ of \cite{Skvortsov:2022syz}.}
\begin{align}
    \mathcal{V}(f,g)&= \exp{[p_{01}+p_{02}]}f(y_1)\wedge g(y_2)\Big|_{y_i=0}\,.
\end{align}
Therefore, the higher spin algebra is  $\hs = \mathrm{Mat}_N\otimes A_1\otimes \mathbb{C}[y]$. The boundary conditions \ref{eq:boundaryconditionsBA}:
\begin{equation}
    \mathcal{V}(e,\omega)+\mathcal{V}(\omega,e)=2\, e^{BB'}y_B\partial_{B'}\omega
\end{equation}
are satisfied with factor $2$.

\paragraph{Coadjoint module.} Similarly, the formal consistency implies that $U(\bullet,\bullet)$ defines a bimodule of $\hs$. The structure maps split as
\begin{align}
    \mathcal{U}(\omega,C)&=\mathcal{U}_1(\omega,C)+\mathcal{U}_2(C,\omega)\,.
\end{align}
The $A_\infty$-relations imply
\begin{equation}
    \begin{split}
        &\mathcal{U}_1(\mathcal{V}(a,b),C)-\mathcal{U}_1(a,\mathcal{U}_1(b,C))=0\,,\\
        &\mathcal{U}_2(\mathcal{U}_1(a,C),b)-\mathcal{U}_1(a,\mathcal{U}_2(C,b))=0\,,\\
        &\mathcal{U}_2(\mathcal{U}_2(C,a),b)+\mathcal{U}_2(C,\mathcal{V}(a,b))=0\,,
    \end{split}
\end{equation}
and can be solved as
\begin{equation}
    \begin{split}
        &\mathcal{U}_1(\omega,C)=+\exp{[p_{02}+p_{12}]}\, \omega(y_1) C(y_2)\Big|_{y_i=0}\,,\\
        &\mathcal{U}_2(C,\omega)=-\exp{[p_{01}-p_{12}]}\, C(y_1) \omega(y_2)\Big|_{y_i=0}\,.
    \end{split}
\end{equation}
The boundary condition \eqref{eq:boundaryconditionsBB} is satisfied with coefficient $2$. As discussed in \cite{Skvortsov:2022syz}, the action of $\mathbb{C}[y]$ in the module is that of differential operators, i.e. $(\omega\circ C)(y)=\omega(\pl) C(y)$, and can be interpreted as the coadjoint action. Indeed, we can define a non-degenerate pairing
\begin{align}
    \langle f| C\rangle&= \exp[p_{12}]f(y_1)C(y_2) \big|_{y_i=0}
\end{align}
between an element $f$ of the algebra $A_1$ and an element $C$ of the dual module $A_0$. In this way we find that $\langle fg| C\rangle=\langle f|\, \mathcal{U}_1(g,C)\rangle$ and $\langle fg| C\rangle=-\langle g|\, \mathcal{U}_2(C,f)\rangle$. Note that formally $\langle f| C\rangle=\langle C| f(-y)\rangle$, which will be useful later to relate $\mathcal{U}$-vertices to $\mathcal{V}$-vertices.

\paragraph{Cubic Vertex $\boldsymbol{\mathcal{V}(\omega,\omega,C)}$.} Cubic vertex, as an $A_\infty$-map, splits into three components
\begin{align}
    \mathcal{V}(\omega,\omega,C)=\mathcal{V}_1(\omega,\omega,C)+\mathcal{V}_2(\omega,C,\omega)+\mathcal{V}_3(C,\omega,\omega)    \,.
\end{align}
The consistency conditions are self-evident and can be found in \cite{Skvortsov:2022syz}. The most interesting fact is that there is a nontrivial solution \cite{Skvortsov:2022syz}
\besubeqs
\begin{align}
    \mathcal{V}_1(\omega,\omega,C)&: && +p_{12} \int_{\Delta_2}\exp[\left(1-t_1\right) p_{01}+\left(1-t_2\right) p_{02}+t_1 p_{13}+t_2 p_{23}]\,, \\
    \mathcal{V}_2(\omega,C,\omega)&: &&\begin{aligned}
       -p_{13}& \int_{\Delta_2}\exp[\left(1-t_2\right) p_{01}+\left(1-t_1\right) p_{03}+t_2 p_{12}-t_1 p_{23}]+\\
       -&p_{13} \int_{\Delta_2}\exp[\left(1-t_1\right) p_{01}+\left(1-t_2\right) p_{03}+t_1 p_{12}-t_2 p_{23}]\,,
    \end{aligned}
    \\
    \mathcal{V}_3(C,\omega,\omega)&: && +p_{23} \int_{\Delta_2}\exp[\left(1-t_2\right) p_{02}+\left(1-t_1\right) p_{03}-t_2 p_{12}-t_1 p_{13}]\,.
\end{align}
\esubeqs
Here, $\Delta_n$ is an $n$-dimensional simplex $t_0=0\leq t_1\leq ...\leq t_n\leq 1$. 

\paragraph{Cubic Vertex $\boldsymbol{\mathcal{U}(\omega,C,C)}$. } Similarly, we can treat $\mathcal{U}(\omega,C,C)$, which is another set of trilinear $A_\infty$-maps now taking values in the module:
\begin{align}
    \mathcal{U}(\omega,C,C)=\mathcal{U}_1(\omega,C,C)+\mathcal{U}_2(C,\omega,C)+\mathcal{U}_3(C,C,\omega)   \,. 
\end{align}
A nontrivial solution reads \cite{Skvortsov:2022syz}:
\besubeqs
\begin{align}
    \mathcal{U}_1(\omega,C,C)&: && +p_{01} \int_{\Delta_2}\exp[\left(1-t_2\right) p_{02}+t_2 p_{03}+\left(1-t_1\right) p_{12}+t_1 p_{13}]\,, \\
    \mathcal{U}_2(C,\omega,C)&: &&\begin{aligned}
       -p_{02}& \int_{\Delta_2}\exp[t_2 p_{01}+\left(1-t_2\right) p_{03}-t_1 p_{12}+\left(1-t_1\right) p_{23}]+\\
       -&p_{02} \int_{\Delta_2}\exp[t_1 p_{01}+\left(1-t_1\right) p_{03}-t_2 p_{12}+\left(1-t_2\right) p_{23}]\,,
    \end{aligned}
    \\
    \mathcal{U}_3(C,C,\omega)&: && +p_{03} \int_{\Delta_2}\exp[\left(1-t_1\right) p_{01}+t_1 p_{02}+\left(t_2-1\right) p_{13}-t_2 p_{23}]\,.
\end{align}
\esubeqs
The trilinear $A_\infty$-maps were found by brute-force in \cite{Skvortsov:2022syz}. This strategy is efficient for the lowest order deformation, but becomes less enjoyable for higher orders. Therefore, we resort to other tools. It is worth noting here, that the action of Chiral Theory is cubic in the light-cone gauge, i.e. on a very specific background. General backgrounds and covariantization may require more contact terms. It is easy to see that the FDA of \cite{Skvortsov:2022syz} has to be completed with higher order structure maps, which is the main goal of this paper. Another fact that is hard not to notice and calls for an explanation is that all $\mathcal{V}$-vertices look like permutations of a single basis structure, the same is true about $\mathcal{U}$-vertices. Moreover, $\mathcal{U}$'s are very similar to $\mathcal{V}$'s.

\subsection{Higher orders}
\label{sec:higherorders}
Since Chiral Theory is local, there must be certain restrictions imposed on the structure maps containing more than one $C$. Indeed, $C$ encodes arbitrarily high derivatives of the dynamical fields. Therefore, any harmlessly looking $\mathcal{V}(...,C,...,C)$ can hide an infinite sum in derivatives. If we choose $\mathcal{V}(\omega,\omega, C,...,C)$ and $\mathcal{U}(\omega, C,...,C)$ for concreteness, the locality forbids infinite tails in $p_{ij}$ with $2<i,j$ for the first and $p_{ij}$ with $1<i,j$ for the second.\footnote{We should recall that there is a silent star-product over $\bry$'s. Therefore, all $r_{ij}$'s are present in the $\exp{[...]}$. Each pair of contracted derivatives on two fields $...\pl_{AA'}\bullet \pl^{AA'}\bullet$ comes from $r_{ij} p_{ij}$. Given that $r_{ij}$ is already present we have to forbid infinite tails in some $p_{ij}$'s. } For example, $p_{23}$ is not found in $\mathcal{U}_1(\omega,C,C)$ at all (having it in the prefactor would still be admissible).

\paragraph{Ingredients of the Homological Perturbation Theory soup.} The main idea is to use the technique of multiplicative resolutions and homological perturbation theory,\footnote{This is a refinement of the original idea of \cite{Vasiliev:1990cm,Vasiliev:1990vu} to introduce additional variables $z$ as to enlarge the field content and write down certain simple equations constraining the $z$-dependence in such a way that perturbative solution to the $z$-equations would reproduce the sought-for vertices. Since all (hypothetical) $4d$ HiSGRA can be cast into the FDA form and, hence, have the same $\omega$ and $C$ field content, we will try to use a notation designed to reveal similarity to \cite{Vasiliev:1990cm,Vasiliev:1990vu}, stressing the important differences along the way. } see \cite{Sharapov:2017lxr,Sharapov:2018hnl,Sharapov:2018ioy,Li:2018rnc} for the discussion and applications to higher spin gravity motivated problems. In a few words, one can try to look for an embedding of the Hochschild complex into a bigger bicomplex, where one can apply different spectral sequences to get explicit formulas for the cocycles. Higher order corrections can also be obtained via homological perturbation theory. 

A suitable  resolution is obtained by extending the commutative algebra $\mathbb{C}[y]$ to another commutative algebra $\mathbb{C}[y,z]$ of polynomials in  $y_{A}$ and $z_{A}$ equipped with a peculiar product. Assuming momentarily we are dealing with deformation quantization type problems, a quite general class of star-products on functions $\mathbb{C}[Y]$ of $Y\equiv Y^a$, $a=1,...,2n$ is determined by matrices $\Omega^{ab}$ via 
\begin{align}\label{stpr}
    (f\star g)(Y)&= \exp{[Y^a \pl^1_a +Y^a\pl^2_a +\Omega^{ab} \pl^1_a \pl^2_b]} \, f(Y_1)g(Y_2)\big|_{Y_{1,2}=0}\,.
\end{align}
The symplectic structure $C^{ab}$ is given by the anti-symmetric part of $\Omega^{ab}$ and the symmetric part is responsible for the choice of ordering (e.g. normal, anti-normal, totally-symmetric or Weyl). One can also write down an integral representation of the same star-product, which is sometimes more useful,\footnote{Here and in what follows we assume that the integrals are defined in such a way that $\int \exp[u^a\xi_a] du=\delta (\xi)$. This is the only formula that we will need to use. }
\begin{align}
    (f\star g)(Y)&=\int dU\,dV\, d\xi\,d\eta\,f(Y+U)g(Y+V) \exp{[U^m\xi_m+V^m\eta_m -\Omega^{mn}\xi_m \eta_n]}\,.
\end{align}
We choose $\Omega^{ab}$ to be symmetric matrix of the form  
\begin{align}
    \Omega=-\begin{pmatrix}
        0 & \epsilon \\
        -\epsilon & 0
    \end{pmatrix}\,,\qquad \epsilon\equiv \epsilon^{AB}\,.
\end{align}
Via the usual formulas it defines a commutative product, whose differential form has symbol
\begin{align}
    \exp{[ p_{01}+p_{02}+q_{01}+q_{02} +p_1\cdot q_2-q_1\cdot p_2 ]}\,,
\end{align}
where we defined $z^{A}\equiv q_0^{A}$, $\pl^{z_i}_{A}\equiv q_{A}^i$. The integral representation reads 
\begin{align}
    (f\star g)(y,z)&= \int du\,dv\,dp\,dq\,f(y+u,z+v) g(y+q,z+p) \exp{[v\cdot q-u\cdot p]}\,.
\end{align}
We will sometimes use notation $\mu(f,g)\equiv f\star g$. The generators $y$ and $z$ act as follows:
\begin{align*}
    y_{A}\star f &= f \star  y_{A}=  (y_{A}  -\pl^z_{A})f\,,   &  z_{A}\star f&=f\star z_{A}= (z_{A}+\pl^y_{A}) f\,.
\end{align*}
Despite commutativity, the star-product has some other interesting properties, the most important being the existence of the element $\varkappa = \exp[z^{C}y_{C}]$ satisfying the relations $$y_{A} \star \varkappa = \varkappa \star y_{A} = z_{A}\star\varkappa =\varkappa\star z_{A}=0\,.$$ We will consider an extension of algebra $A$ by the algebra of differential forms in $d z$ with exterior differential $d_z$ and a familiar two-form $\lambda= \tfrac12 \varkappa\, d z^2$.
We will also repeatedly use the Poincare lemma in the form of
\begin{align}\label{homofor}
      f^{(1)}=h[f^{(2)}]&= d z^{A}\, z_{A} \int_0^1 t\, dt\, f^{(2)}(tz)\,, &
      f^{(0)}=h[f^{(1)}]&= z^{A} \int_0^1 dt\, f_{A}^{(1)}(tz)\,,
\end{align}
see e.g. \cite{Didenko:2014dwa}. 
The first part gives a particular solution to $d_z f^{(1)}= f^{(2)}$ for a one-form $f^{(1)}\equiv dz^{A} f^{(1)}_{A}(z)$ and a given two-form $f^{(2)}\equiv \tfrac12 f^{(2)}(z) \epsilon_{AB}dz^{A} dz^{B}$. The second part gives a particular solution to $d_z f^{(0)}= f^{(1)}$ for a closed one-form $f^{(1)}$ and a zero-form $f^{(0)}\equiv f^{(0)}(z)$. We also complete this definition with $h[f^{(0)}]=0$ for any zero-form $f^{(0)}$.

With the definitions above we are ready to present the whole set of $\mathcal{U}$ and $\mathcal{V}$ vertices defining Chiral HiSGRA (\ref{eq:chiraltheory}). Both type of vertices are constructed as compositions of only two operations: the contracting homotopy $h$ and the $\star$-product (\ref{stpr}); the latter will be also denote by $\mu$. Suitable compositions are conveniently depicted by directed tree graphs, which consist of trivalent vertices, internal edges,  and external edges.  Both ends of an internal edge are on two vertices. Each vertex has two incoming and one outgoing edges. An external edge has one end on a vertex and another end is free. The graphs are supposed to be connected. All the vertices correspond to the star product $\mu$, while the internal edges depict the action of the contracting homotopy $h$:

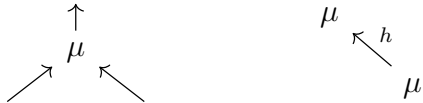
$$ \begin{tikzcd}[column sep=small,row sep=small]
   & {}& \\
    & \mu\arrow[u]  & \\
    \arrow[ur]  & & \arrow[ul]   & 
\end{tikzcd} \;\;\;\;\;\;\;\;\;\;\;\;
\begin{tikzcd}[column sep=small,row sep=small]
    \mu & \\
    &\mu\arrow[ul, "h" ']   
\end{tikzcd}
$$
The external incoming edges (or leaves) correspond to the arguments $\omega$ and $C$ of the interaction vertices $\mathcal{V}$ and $\mathcal{U}$. Therefore, each graph may be decorated with either one or two $\omega$'s, depending on the type of a vertex. As to the  arguments $C$, all of them enter the interaction vertices through a special combination $\Lambda[C]= h[C\diamond\lambda]$, where $C=C(y)$ and $\diamond$ is defined as  $$C\diamond g(z,y)\equiv g(z, y+p_i) \,C(y_i)\,.$$ 
The expression $\Lambda[C]$ decorates the remaining leaves.
Finally, the only outgoing edge (or root) of a connected tree corresponds to the value of an interaction  vertex. 
The $\star$-product being commutative, many trees lead to analytical expressions  that differ only by a permutation of arguments. One should keep in mind that $h^2=0$ and forms of degree higher than two vanish identically. There are also certain classes of trees that vanish due to specific properties of the resolution.\footnote{At this point we can list the crucial similarities/differences as compared to \cite{Vasiliev:1990cm,Vasiliev:1990vu}. The spectrum of the fields (coordinates on $\mathcal{N}$) is exactly the same. The higher spin algebras are different: star-product in $y,\bry$ as compared to star-product in $\bry$ and commutative algebra in $y$. This entails the second difference: zero-forms are no longer in the twisted-adjoint representation \cite{Vasiliev:1999ba}, but in the coadjoint, and form a genuine module of the higher spin algebra. The vertices are all, of course, different. The ones in the present paper are local, those of \cite{Vasiliev:1990cm,Vasiliev:1990vu} form a gauge-invariant ansatz where infinitely many copies of the same interaction are present in different forms (with higher and higher derivatives), field redefinitions are not fixed and, as a result, there are infinitely many free parameters hidden. For generic choice of these parameters, e.g. the one made in \cite{Vasiliev:1990cm,Vasiliev:1990vu}, one gets nonsensical results for correlation functions \cite{Boulanger:2015ova}, which is to be expected and has little to do with the higher spin problem. The whole class of theories sought for in \cite{Vasiliev:1990cm,Vasiliev:1990vu} cannot be constructed with the help of the standard field theory tools due to the non-locality \cite{Dempster:2012vw,Bekaert:2015tva,Maldacena:2015iua,Sleight:2017pcz,Ponomarev:2017qab}, which makes it an interesting challenge to find more general principles to deal with field theories with such non-localities. The resolution, i.e. the $z$-extension we use is also different from \cite{Vasiliev:1990cm,Vasiliev:1990vu}, even though the number of $z$-variables is the same (ignoring $\brz$ that we do not need). At the most basic level the integral form of $\mu$-product contains two integrals more because matrix $\Omega^{ab}$ has rank four, while in \cite{Vasiliev:1990cm,Vasiliev:1990vu} it has rank two. As a result, there is a smooth deformation of our $\mu$-product that leads to \cite{Vasiliev:1990cm,Vasiliev:1990vu}, but not the other way around. }  All admissible trees that contribute to the interaction vertices are generated via Homological Perturbation Theory, which is detailed in Appendices \ref{app:hpt} and \ref{app:trees}. It might be well to point out that the resulting analytical expressions for the vertices $\mathcal{V}$ and $\mathcal{U}$ do not depend on $z$'s as it must if one treats them 
as elements of the higher spin algebra $\hs$ and its coadjoint module, respectively.  Below we present some explicit expressions, starting with the already known results at NLO.

\paragraph{NLO.} The product being commutative, to the NLO we find a single graph to evaluate up to obvious permutations of arguments:
$$
   a(y) \star h[ b(y) \star \Lambda[c(y)] ]= \begin{tikzcd}[column sep=small,row sep=small]
   & {}& \\
    & \mu\arrow[u]  & \\
    a\arrow[ur]  & & \mu\arrow[ul, "h" ']   & \\
    & b\arrow[ur]& &\Lambda[c]\arrow[ul]
\end{tikzcd}
$$
The analytical expression for this graph leads exactly to $\mathcal{V}_1(a,b,c)$, $a,b\in A_1$ and $c\in A_0$. The other $\mathcal{V}$-vertices are simple permutations of this expression. As an illustration let us evaluate the graph step by step in terms of symbols:
\begin{align*}
    c\diamond \lambda&= \tfrac12 \varkappa(z, y+p_3)\, dz^2 \, c(y_3)\,, & \Lambda&= dz^{A}z_{A} \int_0^1 t\,dt\, \varkappa(t z, y+p_3)\, c(y_3)\,,
\end{align*}
we go up the tree to evaluate the product
\begin{align*}
    b(y)\star \Lambda&= dz^{A}(z_{A}+p^2_{A})\, e^{y p_2} \int_0^1 t\,dt\, \varkappa(t z+t p_2, y+p_3)\, b(y_2) c(y_3)\,,
\end{align*}
the resulting expression is still a one-form and $h$ of it is
\begin{align*}
    h[b(y)\star \Lambda]&= (z\cdot p_2)\, e^{y p_2}\int_0^1 dt'\, t\,dt\, \varkappa(tt' z+t p_2, y+p_3)\, b(y_2) c(y_3)\,.
\end{align*}
Lastly, we evaluate one more product and set $z=0$ to find\footnote{HPT ensures that the final answer is $z$-independent, hence, we can set $z=0$ or to any other value.}
\begin{align*}
   a(y)\star h[b(y)\star \Lambda]&= (p_1\cdot p_2)\, e^{y p_1+y p_2}\int_0^1 dt'\, t\,dt\, \varkappa(tt' p_1+t p_2, y+p_3)\, a(y_1) b(y_2) c(y_3)\,.
\end{align*}
We rename $y\rightarrow p_0$ and change the integration variables to those of the $2d$ simplex $\Delta_2$, $t_1=tt'$, $t_2=t$, to arrive at
\begin{align*}
   \mathcal{V}(a,b,c)&= p_{12}\, e^{p_{01}+p_{02}} \int_{\Delta_2} \varkappa(t_1 p_1+t_2 p_2, p_0+p_3)\, a(y_1) b(y_2) c(y_3)\big|_{y_i=0}\,,
\end{align*}
which coincides with $\mathcal{V}_1(a,b,c)$. The other vertices correspond to various permutations of the leaves of the same graph. 

An interesting observation is that we do not have to compute $\mathcal{U}$-vertices separately. All of them can be obtained via duality. Indeed, our $A_\infty$-algebra is very special in that the module $A_0$ is the dual of the algebra $A_1$, $A_0=A_1^*$. With coordinates of the latter denoted (by abuse of notation) by $\omega^\aA$ and coordinates of the former denoted by $C_\aA$, $\mathcal{V}$-vertices correspond to structure constants of type $\mathcal{V}\fudu{\aA}{\aB\aC}{\aD_1...\aD_{n-2}} \omega^\aB \omega^\aC C_{\aD_1}...C_{\aD_{n-2}}$, while $\mathcal{U}$-vertices correspond to $\mathcal{U}\fdu{\aA\aB}{\aD_1..\aD_{n-1}}\omega^\aB C_{\aD_1}...C_{\aD_{n-1}}$. Therefore, we can pair $\mathcal{V}^\aA$ with $C_\aA$ and remove one $\omega$ to get $\mathcal{U}$-vertex (one should keep track of the order of the arguments, of course). For example, we pair $f(p_0,...,p_3)=\langle u |\mathcal{V}_1(a,b,v)\rangle $, where $a,b\in A_1$ and $u,v\in A_0$, which is equivalent to redefining $p_0\equiv \pl^y$. Next, we cyclically rotate $f(-p_3,p_0,p_1,p_2)$ to get $\mathcal{U}_1(p_0,p_1,p_2,p_3)$. 

\paragraph{NNLO.} A crucial test of the approach proposed in the paper is to check the vertices that contain quartic interactions from the action vantage point. Thanks to the symmetries and certain other identities there are only two graphs to evaluate, which we choose to be
$$
   G_1=a(y) \star h[ h[ b(y) \star \Lambda[u] ] \star \Lambda[v]]= \begin{tikzcd}[column sep=small,row sep=small]
   &{}&\\
    & \arrow[u] \mu  & \\
    a\arrow[ur]&&  \arrow[ul,"h"']\mu &  \\
    & \arrow[ur,"h"]\mu &&\arrow[ul]\Lambda[v] \\
    \arrow[ur]b & & \arrow[ul]\Lambda[u] &
\end{tikzcd}
$$
and
$$
   G_2= h[ h[ a(y) \star \Lambda[u] ]\star h[ h[ b(y) \star \Lambda[v] ]= \begin{tikzcd}[column sep=small,row sep=small]
   &&&{}&&&\\
    &&& \arrow[u]\mu  &&& \\
    & \mu\arrow[urr,"h"]& && & \arrow[ull,"h"']\mu &  \\
    a\arrow[ur]&& \arrow[ul]\Lambda[u]  & &  b\arrow[ur]&&\arrow[ul]\Lambda[v]
\end{tikzcd}
$$
The explicit expressions for $G_1$ and $G_2$ can be found in Appendix \ref{app:nnlo}. Below, we highlight some important properties. The general structure is
\begin{align}
    G_1&= \ast (p_{12})^2\exp{[ \ast p_{01}+ \ast p_{02} +\ast p_{13} +\ast p_{23} +\ast  p_{14} +\ast  p_{24} ]}\,,\\
    G_2&=\ast(p_{13})^2 \exp{[\ast p_{01} +\ast p_{03} +\ast p_{12} +\ast p_{23} +\ast p_{14} +\ast p_{34}]}\,,
\end{align}
where $\ast$ denotes certain simple functions of the four integration variables. The integrals, of course, converge. The locality of the vertices is also manifest: $G_1$ does not contain $p_{34}$, while $G_2$ does not have $p_{24}$. Therefore, infinite tails in derivatives are not present. The vertices are obtained by taking permutation of these two graphs and putting them together whenever possible. For example, 
\begin{align}
    \mathcal{V}(\omega,\omega,C,C)&= G_1(\omega,\omega,C,C)\,.
\end{align}
In Appendix \ref{app:tests}, we collected explicit formulas for the quartic vertices in terms of $G_1$, $G_2$ and we also checked several of the consistency relations, i.e. the $A_\infty$-relations, to the lowest order as to make sure that the normalization of the graphs is correct. The $\mathcal{U}$-vertices can be obtained via duality. This completes the story of the quartic vertices (quartic in the equations of motion).

\paragraph{Higher orders.} An interesting property of the resolution proposed in this paper is that the perturbative expansion never terminates. On the contrary, the light-cone action terminates at cubic interactions. There is no tension between these two facts since covariantization of the light-cone action may require infinitely many contact terms. On one hand, the vertices found in \cite{Skvortsov:2022syz} reveal certain minimality. On the other hand, we may still be unlucky in not finding the most minimal field frame where one observes a finite number of nontrivial $A_\infty$ structure maps. It would be interesting to scrutinize this issue more. 

Several comments can be made about higher order vertices without having to compute them explicitly. In Appendix \ref{app:trees}, we describe the class of nontrivial trees that survive the homological perturbation theory. It can also be shown that all higher order vertices are local, i.e. there is a finite number of derivatives provided all helicities are fixed. This is sufficient to prove that the FDA of Chiral Theory is found. For example, just for fun, we computed the quintic vertex in Appendix \ref{app:quintic}. Nevertheless, certain aesthetical aspects are not yet satisfactory, e.g. starting from NNLO the vertices contain quite complicated integrals. It well may be that there are resolutions\footnote{It is determined by the choice of (a) an extension of $y$; (b) an ordering in the most simple case; (c) a contracting homotopy which may differ from \eqref{homofor}.} that lead to simpler results, see e.g. \cite{Iazeolla:2017dxc,DeFilippi:2021xon} for the discussion of the effects of ordering. It may also be useful for applications to fold the homological perturbation theory and rewrite it as certain simple equations, as it was done in \cite{Vasiliev:1990cm,Vasiliev:1990vu}, the danger being that the equations may admit formal solutions that lead to ill-defined vertices from the field theory point of view. 

\section{Conclusions and discussion}
\label{sec:}
The main result of the present paper is the first example of a manifestly Lorentz covariant Higher Spin Gravity with propagating massless fields,\footnote{Other models mentioned in the Introduction are either topological or feature conformal fields. Recent interesting ideas include \cite{deMelloKoch:2018ivk,Aharony:2020omh} and \cite{Sperling:2017dts,Tran:2021ukl,Steinacker:2022jjv}.} which is to be compared to the light-cone action of \cite{Metsaev:1991mt,Metsaev:1991nb,Ponomarev:2016lrm}. The price to pay at the moment is that we have the minimal model rather than an explicit Lorentz-invariant action. The minimal model immediately gives the classical equations of motion as a Free Differential Algebra, but to extract more information one has to compute the $Q$-cohomology at the very least.  

Let us compare the light-cone action of Chiral Theory with the perturbative expansion of its FDA formulation. It is convenient to `pack' all negative helicity fields and the scalar into $\Psi$ and all positive helicity fields into $\Phi$. The light-cone action very schematically reads
\begin{align}\label{sketch}
    \mathcal{L}&= \Psi \square \Phi + c_{+++}\Phi\Phi\Phi+c_{++-}\Phi\Phi\Psi+c_{+--}\Phi\Psi\Psi
\end{align}
and leads to the following, schematic as well, equations of motion:
\begin{align}
    \square \Phi&= c_{++-}\Phi\Phi +c_{+--}\Phi\Psi\,, & \square\Psi&=c_{+++}\Phi\Phi+c_{++-}\Phi\Psi +c_{+--}\Psi\Psi\,.
\end{align}
Assuming that the flat space corresponds to some $\omega_0$ and $C=0$ solution, we would like to compare these equations with 
\besubeqs\label{eq:chiraltheoryA}
\begin{align} 
    D\omega&= \mathcal{V}(\omega, \omega) +\mathcal{V}(\omega_0,\omega,C)+\mathcal{V}(\omega_0,\omega_0,C,C)+...\,,\\
    DC&= \mathcal{U}(\omega,C)+ \mathcal{U}(\omega_0,C,C)+... \,.
\end{align}
\esubeqs
Here $D= d-\omega_0$ is the background covariant derivative in the appropriate representations of the higher spin algebra.
It is clear that only the vertices explicitly indicated above may appear. Indeed, the one-form $\omega_0$, being a purely spin-two background,  is bilinear in $y$, $\bry$. On the other hand, higher vertices involve higher powers of $p_{ij}$ as overall factors acting on $\omega_0$. Therefore, they have to vanish. An important property of the FDA is that all the vertices are local: if we fix helicities of all fields entering a given vertex, then there is only a finite number of derivative involved.

It would be interesting to understand the algebraic side of the story presented in this paper. On one hand, the nontrivial algebraic structures originate from the commutative algebra of polynomials $\mathbb{C}[y]$. On the other hand, the presence of the other tensor factors, the first Weyl algebra $A_1$ for $\bry$ and/or matrix factors, is important for the the whole deformation to be nontrivial. The specific tensor factor of $A_1$ also provides an interpretation to the FDA as a certain $4d$ gauge field theory. This leads to a twist that we can replace $A_1$ by any associative and non-commutative algebra. In particular, this 
may give some new integrable theories in $2d$. Following the general ideas developed in \cite{Sharapov:2018kjz,Sharapov:2019vyd,Sharapov:2020quq} it would be interesting to compute the relevant Chevalley--Eilenberg cohomology to see how unique the above FDA is. The results of the paper can be generalized, with the help of \cite{Sezgin:2012ag}, to the supersymmetric extensions of Chiral Theory and its contractions \cite{Devchand:1996gv,Metsaev:2019dqt,Metsaev:2019aig}.

Since the FDA of the present paper gives the minimal model of Chiral HiSGRA, it contains the same information as the BV-BRST formulation of this theory. Therefore, one can address various questions (actions, counterterms, anomalies, etc.). For example, Chiral Theory was shown to be one-loop finite \cite{Skvortsov:2018jea,Skvortsov:2020wtf,Skvortsov:2020gpn}, but extending these results to higher orders is challenging in the light-cone gauge. It would also be interesting to construct exact solutions, which should more easily result from a twistor formulation of Chiral HiSGRA that is yet to be found, see \cite{Tran:2021ukl} for the first steps in this direction. The results of the present paper should also be helpful in looking for the twistor action of Chiral Theory. Lastly, the generalization of the results of this paper to anti-de Sitter space is, in principle, straightforward\footnote{In this regard, it is worth stressing that (contrary to the old higher spin folklore that there cannot be smooth limit) Chiral Theory has a smooth deformation to $(A)dS_4$ or, equivalently, the smooth flat limit, \cite{Metsaev:2018xip,Skvortsov:2018uru}.} \cite{Sharapov:2022awp} and has important implications for Chern-Simons Matter theories and $3d$ bosonization duality. Nevertheless, the exact reasons for why the ideas of the present paper and the extension thereof to $(A)dS_4$ \cite{Sharapov:2022awp} work, i.e. give a local HiSGRA in one step, require better understanding. It should also be kept in mind that the hypothetical twistor action would automatically lead to Minkowski and (anti)-de Sitter versions of Chiral Theory depending on the infinity twistor chosen.

\section*{Acknowledgments}
\label{sec:Aknowledgements}
The work of E.S. and R.van D was partially supported by the European Research Council (ERC) under the European Union’s Horizon 2020 research and innovation programme (grant agreement No 101002551) and by the Fonds de la Recherche Scientifique --- FNRS under Grant No. F.4544.21. A. Sh. gratefully acknowledges the financial support of the Foundation for the Advancement of Theoretical Physics and Mathematics “BASIS”.  The results of Appendix B on HPT for higher spin algebras were obtained under exclusive support of the Ministry of Science and Higher Education of the Russian Federation (project No. FSWM-2020-0033). The work of A.Su. was supported by the Russian Science Foundation grant 18-72-10123 in association with the Lebedev Physical Institute.

\appendix

\section{NNLO vertices}
\label{app:nnlo}
At NNLO the two graphs that contribute (up to permutations and duality) can be evaluated to give
\begin{align*}
    G_1&=p_{12}^2 \int_0^1dt_1\int_0^1dt_2\int_0^1dk_1\int_0^1dk_2\times\frac{k_1t_1t_2\left(1-t_1\right)\left(1-t_2\right)}{\left(1-t_1k_2t_2\right)^4}\\
    &\times \exp\Big[p_{01}\left(\frac{k_1 \left(k_2 t_1-2 k_2 t_2 t_1+t_2\right)}{k_2 t_1 t_2-1}+1\right)
    -p_{02}\frac{\left(t_1-1\right) \left(t_2-1\right)}{k_2 t_1 t_2-1}+\nonumber\\
    &+p_{13}\frac{k_1 k_2 t_1 \left(t_2-1\right)}{k_2 t_1 t_2-1}+p_{14}\frac{k_1 t_2 \left(k_2 t_1-1\right)}{k_2 t_1 t_2-1}+\nonumber\\
    &+\left.p_{23}\frac{t_1 \left(k_2 t_2-1\right)}{k_2 t_1 t_2-1}+p_{24}\frac{t_2\left(1-t_1\right)}{1-t_1k_2t_2}\right]a\left(y_1\right)b\left(y_2\right)c\left(y_3\right)d\left(y_4\right)\big|_{y_i=0}
\end{align*}
and (we tried to manifest $\Delta_2\times \Delta_2$-structure of the integral)
\begin{align*}
    G_2&=-p_{13}^2 \int_0^1dt_1\int_0^1dt_2\int_0^{t_1}dk_1\int_0^{t_2}dk_2\times\frac{\left(1-t_1\right)\left(1-t_2\right)}{\left(1-k_1k_2\right)^4}\\
    &\times \exp\left[p_{01}\frac{\left(1-t_1\right)\left(1-k_2\right)}{1-k_1k_2}+p_{03}\frac{\left(1-t_2\right)\left(1-k_1\right)}{1-k_1k_2}+\right.\nonumber\\
    &+\left.p_{12}\frac{t_1-k_1k_2}{1-k_1k_2}+p_{34}\frac{t_2-k_1k_2}{1-k_1k_2}+p_{14}\frac{k_2\left(1-t_1\right)}{1-k_1k_2}-p_{23}\frac{k_1\left(1-t_2\right)}{1-k_1k_2}\right]\times\nonumber\\
    \times&a\left(y_1\right)c\left(y_2\right)b\left(y_3\right)d\left(y_4\right)\big|_{y_i=0}
\end{align*}
The integrals are convergent. As discussed in the main text, the graphs give local contributions to the vertices.

\section{Homological perturbation theory}
\label{app:hpt}
In this section, we show how to obtain all the interaction vertices of Chiral HiSGRA by means  of {\it homological perturbation theory} (HPT). A detailed account of the theory can be found in \cite{HK}, \cite{GLS}, \cite{Kr} (see also \cite{Li:2018rnc} for a similar discussion of HPT in the context of formal HiSGRA).

As in the main text, we start with the cochain complex  $A=\mathbb{C}[y, z, dz]$  of differential forms with polynomial coefficients. The coboundary operator $d_{z}: A_{n}\rightarrow A_{n+1}$ is given by the usual 
exterior differential on $z$'s.  Combining the exterior product of the basis differentials $dz^{A}$ with the $\star$-product 
\begin{equation}\label{pp}
a\star b=a(y,z)\exp\Big({\frac{\stackrel{\leftarrow}{\partial}}{\partial z^{A}}\frac{\stackrel{\rightarrow}{\partial}}{\partial y_A}}-\frac{\stackrel{\leftarrow}{\partial}}{\partial y^{A}}\frac{\stackrel{\rightarrow}{\partial}}{\partial z_{A}}\Big) b(y,z)
\end{equation}
of polynomials in the $y$'s and $z$'s, we get a commutative dg-algebra $(A, d_{z})$. 
Actually, the $\star$-product above is equivalent to the conventional (dot) product on polynomials:
$$
a\star b=e^{-\Delta}\left((e^{\Delta}a)\cdot(e^{\Delta}b)\right)\,,\qquad \Delta=\frac{\partial^2}{\partial y^{A}\partial z_{A}}\,.
$$
The dual space $A^\ast$ carries the canonical structure of a graded bimodule over $A$. In particular, $A^\ast_0$ is clearly isomorphic to the space of formal power series $\mathbb{C}[[y,z ]]$. The left/right action of $A$ on $A_0^\ast$ is given by
$$
a\circ m=m\circ a= (e^{\Delta}a)(\partial_{y},\partial_{z},0)m(y, z)\,,\qquad \forall a=a(y, z,dz)\in A,\quad m\in A^\ast_0\,.
$$
Indeed, 
$$
(a\star b)\circ m=e^{-\Delta}\left((e^{\Delta}a)\cdot(e^{\Delta}b)\right)\circ m
$$
$$
=\left((e^{\Delta}a)\cdot(e^{\Delta}b)\right)(\partial_{y},\partial_{z}, 0) m(y,z)=a\circ(b\circ m)\,,
$$
and the same for the right action. The bimodule $A_0^\ast$ is too big and highly reducible. In the following we will deal with its submodule $M=\mathbb{C}[[y]]\subset A^\ast_0$ constituted by formal power series in $y$'s. Extending the action of $d_{z}$ to $M$ by zero, we can think of $M$ as a differential bimodule over $A$. Furthermore, it is convenient to combine the differential bimodule structure into  a single dg-algebra $\mathcal{A}=A\oplus M$ for the following $\ast$-product and differential:
$$
(a, m)\ast (a', m')=(a\star a', a\circ m'+a'\circ m)\,,\qquad d_{z}(a,m)=(d_{z}a, 0)\,.
$$
By definition, the degree of an elements $a$ of $A$ coincides with its form-degree, while all the elements of $M$ have degree $1$. 

In addition to $d_{z}$ we can endow the algebra $\mathcal{A}$ with one more differential $\delta$ of degree $1$. This is defined as   
\begin{equation}\label{dd}
\delta (a,m)=\big(m(-z)e^{z^{A}y_{A}}dz^{B}\wedge dz_{B}, 0\big)\,,\qquad \forall\; (a,m)\in \mathcal{A}\,.
\end{equation}
(Notice the change of the argument in $m$.) It is clear that $\delta^2=0$. The differential $\delta$ will be a derivation of the $\ast$-product above if and only if the following identities hold: 
$$
\delta(a\circ m)=(-1)^{|a|}a\star \delta m\,,\qquad \delta m\circ m'=m\circ \delta m'\,.
$$
The first equality is enough to check only for the generators  $y^{A}$, $z^{A}$'s and $dz^{A}$. We find
$$
y^{A}\star \delta m=(y^{A}-\partial^{A}_{z})m(-z)e^{ {z}^{B} y_{B}}dz^{C}\wedge dz_{C}=e^{y^{B} z_{B}}(-\partial_{z}^{A} m(-z))dz^{C}\wedge dz_{C}=\delta(y^{A}\circ m)\,,
$$
$$
z^{A}\star \delta m=(z^{A}+\partial^{A}_{y})m(-z)e^{z^{B} y_{B}}dz^{C}\wedge dz_{C}=0=\delta(z^{A}\circ m)\,,
$$
$$
dz^{A}\star \delta m=m(-z)e^{ {z}^{B} y_{B}}dz^{A}\wedge dz^{C}\wedge dz_{C}=0=-\delta(dz^{A}\circ m)\,.
$$
The second identity is also satisfied because $\delta m$ is a two-form and $(\delta m)(y, z,0)=0$.

Since the differentials trivially commute to each other, $d_{z}\delta+\delta d_{z}=0$, we can combine them into the total differential $D=d_{z}+\delta$ of degree $1$. Given now  a dg-algebra $(\mathcal{A}, D)$, one can ask about its minimal model. In general, constructing a minimal model of a dg-algebra is quite  a difficult  problem. What helps us a lot are two things: $(i)$ the differential $d_{z }$, being the exterior differential on polynomial forms, admits an explicit contracting homotopy $h$  and $(ii)$ one may regard $D$ as a `small perturbation' of $d_{z}$ by $\delta$. Under these circumstances, homological perturbation theory offers the most efficient way to build the minimal model in question. As we will see, this minimal model yields  exactly the $A_\infty$-algebra defining the r.h.s. of the field equations in Chiral HiSGRA.
Below we recall some basic definitions and statements. 

\begin{definition}

A {\it strong deformation retract} (SDR) is given by a pair of complexes $(V,d_V)$ and $(W,d_W)$ together with chain maps $p:V\rightarrow W$ and $i:W\rightarrow V$
such that $pi=1_W$ and $ip$ is homotopic to $1_V$. The last property implies the  existence of a map $h: V\rightarrow V$ such that 
$$
dh+hd=ip-1_V\,.
$$
Without loss in generality, one may also assume the following {\it annihilation properties}:  
$$
hi=0\,,\qquad ph=0\,,\qquad h^2=0\,.
$$
\end{definition}
All the data above can be summarized by a single  diagram 
\begin{equation}\label{SDR}
\xymatrix{
*{\hspace{5ex}(V,d_V)\;}\ar@(ul,dl)[]_{h} \ar@<0.5ex>[r]^-p
& (W, d_W) \ar@<0.5ex>[l]^-i}\,.
\end{equation}
Let us mention a special case of this construction where  $W=H(V,d_V)$ is the cohomology group of the complex $(V,d_V)$ and $d_W=0$. 

The main concern of HPT is  transferring various algebraic structures form one object to another through a homotopy equivalence. Whenever applicable, the theory provides effective algorithms and explicit formulas, as distinct from classical homological algebra.  The cornerstone of HPT is 
the following statement, often called the Basic Perturbation Lemma. 
\begin{lemma}[\cite{B}]\label{BPL}
For any SDR data (\ref{SDR}) and a small perturbation $\delta$ of  $d_V$ such that $(d_V+\delta)^2=0$ and $1-\delta h$ is invertible, there is a new SDR 
$$
\xymatrix{
*{\hspace{9ex}(V,d_V+\delta )\;}\ar@(ul,dl)[]_{h'} \ar@<0.5ex>[r]^-{p'}
&(W, d'_W) \ar@<0.5ex>[l]^-{i'}}\,,
$$
where the maps are given by 
$$
\begin{array}{ll}
     p'=p+p(1-\delta h)^{-1}\delta h\,,&\quad i'=i+h(1-\delta h)^{-1}\delta i\,, \\[3mm]
    h'=h+h(1-\delta h)^{-1}\delta h\,, & \quad d'_W = d_W+p(1-\delta h)^{-1}\delta i\,.
\end{array}
$$
\end{lemma}
One can think of the operator $A=(1-\delta h)^{-1}$ as being defined by a geometric series 
\begin{equation}\label{GS}
A=\sum_{n=0}^\infty (\delta h)^n\,. 
\end{equation}
In many practical cases its convergence is ensured by a suitable filtration on $V$. 

We are concerned with transferring  $A_\infty$-structures on $V$ to its cohomology space $W$. 
To put this transference problem into the framework of HPT one first applies the tensor-space functor $T$  to the vector spaces $V$ and $W$. Recall that, in addition to the associative algebra structure, the space $T(V)=\bigoplus_{n\geq 1}V^{\otimes n}$ carries the structure of a coassociative  coalgebra with respect to the coproduct 
$$
\Delta: T(V)\rightarrow T(V)\otimes T(V)\,,
$$
$$
\Delta (v_1\otimes \cdots\otimes v_n)=\sum_{i=1}^{n-1}(v_1\otimes\cdots\otimes v_i)\otimes (v_{i+1}\otimes\cdots\otimes v_n)\,.$$
Coassociativity is expressed by the relation $(1\otimes \Delta)\Delta=(\Delta\otimes 1)\Delta$. A linear map $F: T(V)\rightarrow T(V)$ is called a {\it coderivation}, if it obeys the co-Leibniz rule $$\Delta F=(F\otimes 1+1\otimes F)\Delta\,.$$ 
The space of coderivations is known to be  isomorphic to the space of homomorphisms $\mathrm{Hom}(T(V),V)$, so that  any homomorphism $f: T(V)\rightarrow V$ induces a coderivation $\hat{f}: T(V)\rightarrow T(V)$ and vice versa:  if $f\in \mathrm{Hom}(T^m(V),V)$, then  
\begin{equation}\label{f1}
\begin{array}{rl}
\hat{f}(v_1\otimes\cdots\otimes v_n)=\displaystyle \sum_{i=1}^{n-m+1}&(-1)^{|f|(|v_1|+\cdots+ |v_{i-1}|)}v_1\otimes\cdots\otimes v_{i-1}\\[4mm]
&\otimes f(v_i\otimes \cdots \otimes v_{i+m-1})\otimes v_{i+m}\otimes \cdots\otimes v_n
\end{array}
\end{equation}
for $n\geq m$ and zero otherwise. 

The notion of a coderivation provides an alternative definition of an $A_\infty$-algebra: An $A_\infty$-algebra structure on a graded vector space $V$ is given by an element $m\in \mathrm{Hom}(T(V),V)$ of degree one such that the corresponding  coderivation $\hat m$ squares to zero. Every such $\hat m$ is called a {\it codifferential}.  The condition $\hat m^2=0$ is equivalent to the equation $m\circ m=0$, where $\circ$ stands for the Gerstenhaber product \eqref{gersproduct}. Expanding $m$ into the sum $m=m_1+m_2+\cdots$ of homogeneous multi-linear maps $m_n\in \mathrm{Hom}(T^n(V),V)$ and substituting it back into $m\circ m=0$ gives an infinite sequence of homogeneous  relations on $m$'s, known as Stasheff's identities \cite{St}. In particular, the first structure map $m_1: V_l\rightarrow V_{l+1}$ squares to zero, $m_1^2=0$, making $V$ into a complex of vector spaces.  An  $A_\infty$-algebra is called {\it minimal} if $m_1=0$. For minimal algebras the second structure map  $m_2: V\otimes V\rightarrow V$ makes the space $V[-1]$ into a graded associative algebra with respect to the $\ast$-product\footnote{By definition, $V[-1]_n=V_{n-1}$. On shifting degree by one unit, the $\ast$-product acquires degree $0$.}
\begin{equation}\label{bbb}
a\ast b=(-1)^{|a|}m_2(a\otimes b)\,.
\end{equation}
Associativity is encoded by the Stasheff identity $ m_2\circ m_2=0$. From this perspective, a graded associative algebra is just an $A_\infty$-algebra with $m=m_2$.  More generally, an $A_\infty$-algebra with $m=m_1+m_2$ is equivalent to a  differential graded algebra $(V[-1],\ast, d)$ with the product (\ref{bbb}) and the differential $d=m_1$. Again, the Leibniz rule 
$$
d(a\ast b)=da\ast b+(-1)^{|a|-1}a\ast db
$$
is equivalent to the Stasheff identity $m_1\circ m_2+m_2\circ m_1=0$. 

The next statement, called the {\it tensor trick}, allows one to transfer SDR data from spaces to their tensor (co)algebras. 
\begin{lemma}[\cite{GL}]\label{TT}
With any SDR data (\ref{SDR}) one can associate a new SDR 
$$
\xymatrix{{(\,}\ar@(ul,dl)[]_-{\hat{h}}&
{\hspace{-6ex}T(V),\hat{d}_V )\;} \ar@<0.5ex>[r]^-{\hat{p}}
& (T(W), \hat{d}_W) \ar@<0.5ex>[l]^-{\hat{i}}}\,,
$$ 
where the new differentials $\hat{d}_V$ and $\hat{d}_W$ are defined by the rule (\ref{f1}),
$$
\hat{p}=\sum_{n=1}^\infty p^{\otimes n}\,,\qquad \hat{i}= \sum_{n=1}^\infty i^{\otimes n}\,,
$$
and the new homotopy is given by
$$
\hat{h}=\sum_{n= 1}^\infty \sum_{k=0}^{n-1} 1^{\otimes k}\otimes h\otimes (ip)^{\otimes n-k-1}\,.
$$
\end{lemma}

After reminding the basics of HPT let us return to our deformation problem.
Consider first the case of dg-algebra $\mathcal{A}$ with respect to the unperturbed differential $d_{z}$. 
By the algebraic Poincar\'e Lemma, $H(\mathcal{A}, d_{z})\simeq \mathbb{C}[y]\oplus \mathbb{C}[[y]]$. Here  the first summand corresponds to the differential forms of  $A$ that are independent of $z$'s and $dz$'s, while the second summand is given by the elements of the module $M$. To streamline our notation we will write $\mathcal{H}$ for the algebra of cohomology $H(\mathcal{A}, d_{z})$.  Clearly, the natural inclusion $i :\mathcal{H}\rightarrow \mathcal{A}$ is an algebra homomorphism. This leads us immediately to SDR (\ref{SDR}) with 
$$
V=\mathcal{A}[1] \,,\qquad W=\mathcal{H}[1] \,,\qquad d_V=d_{z}\,, \qquad d_W=0\,,
$$
$$
p (a,m)=(a(y,0,0), m)\,,\qquad h(a, m)=(h(a), 0)\,,
$$
and $h(a)$ was defined in \eqref{homofor} as the standard contracting homotopy for the de Rham complex. 
Applying the tensor trick yields then an SDR for the corresponding tensor (co)algebras
$$
\xymatrix{{(\,}\ar@(ul,dl)[]_-{\hat{h}}&
{\hspace{-6ex} T(\mathcal{A}[1]),\hat{d}_{z})\;} \ar@<0.5ex>[r]^-{\hat{p}}
& (T(\mathcal{H}[1]), 0 ) \ar@<0.5ex>[l]^-{\hat{i}}}\,.
$$
Let $\mu$  denote multiplication (the $\ast$-product) in $\mathcal{A}$. It defines the coderivation $\hat \mu$ such that $(\hat d_{z}+\hat \mu)^2=0$. 
 This allows us to treat $\hat\mu$ as a small perturbation of the differential $\hat{d}_{z}$. By making use of the Basic Perturbation Lemma \ref{BPL}, we  obtain the new SDR 
$$
\xymatrix{{(\,}\ar@(ul,dl)[]_-{\hat{h}'}&
{\hspace{-6ex}T(\mathcal{A}[1]),\hat{d}_{z}+\hat\mu)\;} \ar@<0.5ex>[r]^-{\hat{p}'}
& \big(T(\mathcal{H}[1]), \hat m_2 \big ) \ar@<0.5ex>[l]^-{\hat{i}'}}\,,
$$
where the codifferential on the right is given by 
\begin{equation}\label{hatmt}
  \hat m_2 = \hat{p}(1-\hat\mu \hat{h})^{-1}\hat\mu\hat{i}\,.
\end{equation}
Notice that $\hat h \hat \mu \hat i=0$, because $h$ vanishes on the subalgebra $ i(\mathcal H)\subset \mathcal A$. Hence, 
$\hat m_2=\hat p\hat \mu\hat i$ and the dg-algebra $(\mathcal{A}, d_{z})$ is formal: its minimal model $\mathcal{H}$ involves no higher multiplication operations in addition to the $\ast$-product (\ref{bbb}).

Finally, let us turn to the dg-algebra $(\mathcal{A}, D=d_{z}+\delta)$. This yields the SDR data
$$
\xymatrix{{(\,}\ar@(ul,dl)[]_-{\hat{h}'}&
{\hspace{-6ex}T(\mathcal{A}[1]),\hat{d}_{z}+\hat \delta+\hat\mu)\;}
 \ar@<0.5ex>[r]^-{\hat{p}'}
& \big(T(\mathcal{H}[1]), \hat m \big ) \ar@<0.5ex>[l]^-{\hat{i}'}}\,.
$$
Again, we can regard  the sum $\hat \mu+\hat \delta$ as a small perturbation of the basic differential $\hat d_{z}$. Lemma \ref{BPL} gives then the formal expression for the codifferential $\hat m$ on the right:
\begin{equation}\label{ps}
  \hat m = \hat{p}\big(1-(\hat\mu+\hat \delta) \hat{h}\big)^{-1}(\hat\mu+\hat \delta)\hat{i}=
  \hat m_2+\hat{p}\big(1-(\hat\mu+\hat \delta) \hat{h}\big)^{-1}\hat \delta\hat{i}\,.
\end{equation}
 One can simplify various terms of this formula  by noting that $\hat p\hat \delta=0$ and  $\delta h=0$.  Using (\ref{GS}), one can also find that the deformed codifferential starts as 
$$
\hat m=\hat m_2+\hat p \hat \mu\hat h\hat \mu\hat h\hat \delta \hat i+\cdots\,.
$$
The second term on the right defines the third structure map $m_3$ of the $A_\infty$-algebra $\mathcal{H}[1]$. 
Thus, the dg-algebra $(\mathcal{A}, D)$ is not formal. The diagrams in the main text and Appendices give just a pictorial representation for various terms of the perturbation series (\ref{ps}); in so doing, the inclusion and projection maps $i$ and $p$ correspond to incoming and outgoing edges, respectively. 

\section{Higher orders}
\label{app:trees}
Let us elaborate a bit more on the structure of HPT under consideration. Notice that the image of the differential (\ref{dd}) is not a polynomial function and the $\star$-product (\ref{pp}) of non-polynomial functions in $y$'s and $z$'s is ill-defined.  Therefore, one needs to make sure that the perturbation series (\ref{ps}) does make sense when applied to polynomial functions.

As it was already mentioned, there are many symmetries thanks to the commutativity of the $\star$-product (we denote it $\mu$) that the resolution is based on. The permutation symmetry over the legs attached to $\mu$-vertices is obvious. We would like to show that all nontrivial trees that contribute can be depicted as 
$$
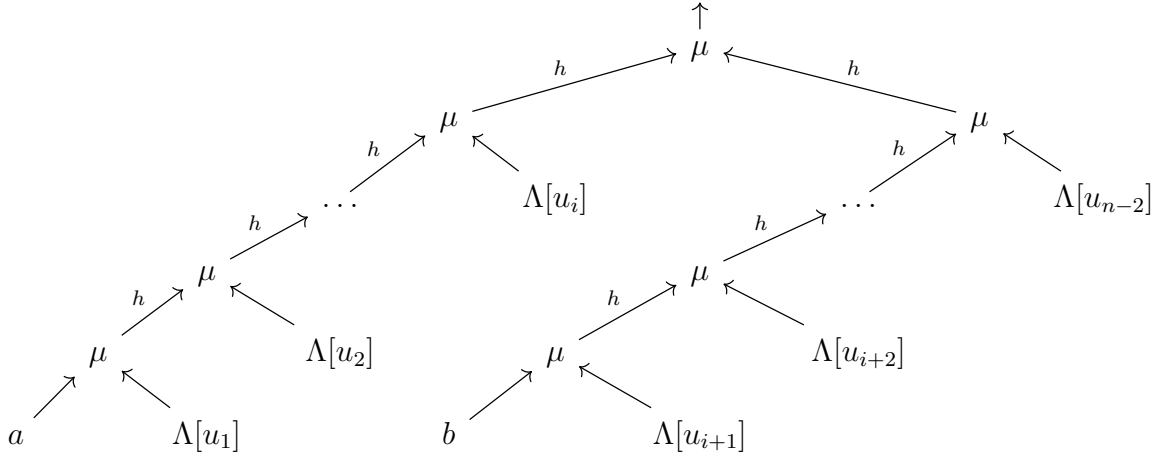
\begin{tikzcd}[column sep=small,row sep=small]
&            & &                   &                           &       & {} &       &               &\\
&            & &                   &                           &       & \arrow[u]\mu &       &               &\\
&            & &                   &\arrow[urr,"h"]\mu    &       &                           &       &\arrow[ull,"h"']\mu&\\
&            & &\dots\arrow[ur,"h"]    &                           &\arrow[ul]\Lambda[u_i]      &                           &\dots\arrow[ur,"h"]  &              &\arrow[ul]\Lambda[u_{n-2}]\\
&           &\mu\arrow[ur,"h"]      &   & & &\mu\arrow[ur,"h"] & &   &\\
&\mu\arrow[ur,"h"] & & \arrow[ul]\Lambda[u_2]                  &                           &       \mu\arrow[ur,"h"]&  &\arrow[ul]\Lambda[u_{i+2}]                 &               &\\
a\arrow[ur]&            &\arrow[ul]\Lambda[u_1]                 & &                           b\arrow[ur]&       &       \arrow[ul]\Lambda[u_{i+1}]& & &                           
\end{tikzcd}
$$
In words, the tree consists of two branches, each having one leaf with an argument from the algebra $A_1$, $a$ or $b$ here. Apart from $a$, or $b$ each of the two branches has only simple leaves with $\Lambda[u_i]=h\delta u_i$, where $u_i$ belong to the module $A_0$. The branches may have different lengths. The graph above is a contribution to the  $A_\infty$-map $m_n$ with $n$ arguments in total:
\begin{align}
    m(a,u_1,\ldots,u_i,b,u_{i+1},\ldots,u_{n-2})\,.
\end{align}
There is a number of simple observations that reduce the variety of trees to the class we described (we introduce one-form $A$ in $z$-space as $A=\Lambda[u_i]$): (i) $h$ cannot be the last operation on a tree since we can set $z=0$ at the end and $h$ has $z$-factor; (ii) $h^2\equiv0$ is obvious; (iii) there are no three-forms, hence, $A\star A\star A \equiv0$; (iv) one can also see that $h(A\star A)\equiv0$; (v) it follows from (i) that the final result should be the product $\mu(T_1,T_2)$ of two sub-trees $T_{1,2}$, each of which being zero-form. In particular, each $T_i$ must have all $A$ balanced by $h$. Given the rules above, it is impossible to construct a zero-form tree only with $A$'s. Therefore, each $T_i$ must have one and only one of the arguments in the algebra, e.g. $a$ belongs to $T_1$ and $b$ belongs to $T_2$. Let us zoom in on one of the two sub-trees. We could see two pictures: \\
\phantom{a}
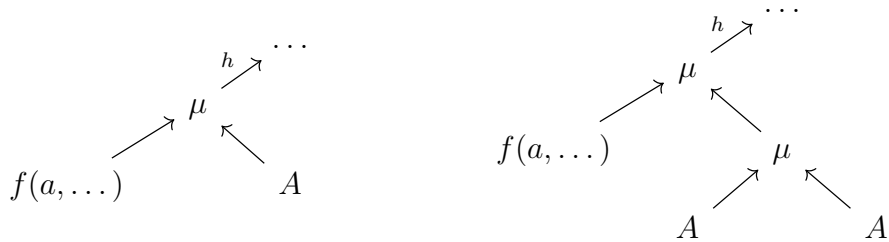
\hspace{2cm}\begin{tikzcd}[column sep=small, row sep=small]
                &                           &\dots\\
                &   \mu\arrow[ur,"h"] & \\
    f(a,\dots)\arrow[ur]  &                           &\arrow[ul]A
\end{tikzcd}\hspace{2cm}
\begin{tikzcd}[column sep=small, row sep=small]
                &                           &\dots           &\\
                &   \mu\arrow[ur,"h"] &                           &\\
    f(a,\dots)\arrow[ur]  &                           &\mu\arrow[ul]    &\\
                &A\arrow[ur]                          &                           &A\arrow[ul]
\end{tikzcd}\\
In fact, the second option is inconsistent. It gives a one-form and we have to find a way to make the whole sub-tree be zero-form at the end. We cannot attach a zero-form sub-tree (and apply $h$ afterwards) since $b$ is in another sub-tree. We can only attach $A$ or any other one-form sub-tree, but this leads to a two-form, i.e. to the original problem we are trying to solve. We are in a vicious circle. Therefore, the second option cannot be realized. $\blacksquare$

\paragraph{Locality.} It is important to prove that the vertices are local in the sense of not having $p_{ij}$ in the exponent that contract some of the zero-form arguments. Given the result above, we can have a look at the general structure of one of the branches. It is easy to see that it has the following general form:
\begin{align}\label{induct}
    h&\left(\cdots h\left(a\star \Lambda[c_2]\right)\dots\star \Lambda[c_n]\right)=\nonumber\\
    &=\eta\left(z p_1\right)^{n-1}\exp\left[\gamma_0z y+\gamma_1y p_1+\gamma_2z p_2+\dots+\gamma_nz p_n+\zeta_2p_{12}+\dots+\zeta_np_{1n}\right]\times\\
    &\times a\left(y_1\right)c_2\left(y_2\right)\cdots c_n\left(y_n\right)\,,\notag
\end{align}
where $\eta$, $\gamma_i$ and $\zeta_i$ are certain functions of the integration variables $t_k$ that originate from multiple applications of $h$. The integral sign is omitted. Indeed, we begin with the lowest possible expression to start the induction
\begin{align*}
    h\left(a\star \Lambda[c]\right)=\int_0^1dt_1\int_0^{t_1}dk_1\left(z p_1\right)\exp\left[y p_1\left(1-t_1\right)+z y k_1+z p_2k_1+p_{12}t_1\right] a\left(y_1\right)c\left(y_2\right)\,.
\end{align*}
Assuming the structure is as in \eqref{induct} we attempt to proceed to the next order to find
\begin{align*}
    &\eqref{induct}\star \Lambda[c_{n+1}]=dz^{A}\,\eta'\left(z p_1\right)^{n-1}\left(\alpha_0z_{A}+\alpha_1p_{A}^1\right)\times\nonumber\\
    \times&\exp\left[\gamma'_0z y+\gamma_1y p_1+\gamma'_2z p_2+\dots+\gamma'_nz p_n+\gamma'_{n+1}z p_{n+1}+\right.\nonumber\\
    +&\left.\zeta'_2p_{12}+\dots+\zeta'_np_{1n}+\zeta'_{n+1}p_{1\ n+1}\right]\times a\left(y_1\right)c_2\left(y_2\right)\cdots c_n\left(y_n\right)c_{n+1}\left(y_{n+1}\right)\,.
\end{align*}
Applying $h$ to the expression here-above we clearly reproduce \eqref{induct}. Now, we can compute the $\mu$-product of two expressions of type \eqref{induct} to see that the final answer has the desired property of being local. $\blacksquare$

\section{Consistency at NNLO}
\label{app:tests}
It is reassuring to check the consistency of the quartic vertices directly, which, in particular, makes sure that the signs/coefficients are correct. We add the quartic term to $d\omega$ and $dC$:
\begin{align*}
    \begin{aligned}
    d\omega&=V(\omega,\omega)+\mathcal{V}_1(\omega,\omega,C)+\mathcal{V}_2(\omega,C,\omega)+\mathcal{V}_3(C,\omega,\omega)+\mathcal{V}_1(\omega,\omega,C,C)+\mathcal{V}_2(\omega,C,\omega,C)\,,\\
    &+\mathcal{V}_3(\omega,C,C,\omega)+\mathcal{V}_4(C,\omega,C,\omega)+\mathcal{V}_5(C,\omega,\omega,C)+\mathcal{V}_6(C,C,\omega,\omega)\\
    dC&=\mathcal{U}_1(\omega,C)+\mathcal{U}_2(C,\omega)+\mathcal{U}_1(\omega,\omega,C)+\mathcal{U}_2(\omega,C,\omega)+\mathcal{U}_3(C,\omega,\omega)+\mathcal{U}_1(\omega,C,C,C)\\
    &+\mathcal{U}_2(C,\omega,C,C)+\mathcal{U}_3(C,C,\omega,C)+\mathcal{U}_4(C,C,C,\omega)\,.
    \end{aligned}
\end{align*}
The consistency condition $0\equiv d^2\omega$ can be split into $10$ equations for different ordering of $\omega\omega\omega C C$. Let us have a look at some of them. We begin with the final answer --- expressions for the vertices in terms of the two graphs $G_1$ and $G_2$ that contribute at the second order:
\begin{align*}
    \mathcal{V}_1(\omega,\omega,C,C)&=G_1(\omega,\omega,C,C)\,,\\
    \mathcal{V}_2(\omega,C,\omega,C)&=-(\sigma_{(423)} G_1)(\omega,C,\omega,C) - (\sigma_{(23)}G_1)(\omega,C,\omega,C) + G_2(\omega,C,\omega,C)\,,\\
    V_3(\omega,C,C,\omega)&=(\sigma_{(24)}G_1)(\omega,C,C,\omega)+(\sigma_{(1432)}G_1)(\omega,C,C,\omega)-(\sigma_{(34)}G_2)(\omega,C,C,\omega)\,,\\
    \mathcal{V}_4(C,\omega,C,\omega)&=-(\sigma_{(124)}G_1)(C,\omega,C,\omega)-(\sigma_{(1243)}G_1)(C,\omega,C,\omega)+(\sigma_{(12)(34)}G_2)(C,\omega,C,\omega)\\
    \mathcal{V}_5(C,\omega,\omega,C)&=-(\sigma_{(12)}G_2)(C,\omega,\omega,C)\,,\\
    \mathcal{V}_6(C,C,\omega,\omega)&=(\sigma_{(14)(23)}G_1)(C,C,\omega,\omega)\,,
\end{align*}
where $\sigma_{(...)...(...)}$ is the standard notation for the decomposition of a given permutation into disjoint cycles. The $\mathcal{U}$-vertices can be obtained via the duality. For example, 
\begin{align*}
    \mathcal{U}_1(\omega,C,C,C)=G_1(\omega,C,C,C)(-p_4,p_0,p_1,p_2,p_3)\,.
\end{align*}

Now let us check directly that some of the $A_\infty$-relations must be satisfied. One of the simplest integrability conditions reads 
\begin{align*}
    -\omega \mathcal{V}_1(\omega,\omega,C,C)+\mathcal{V}_1(\omega^2,\omega,C,C) -\mathcal{V}_1(\omega,\omega^2,C,C)+\mathcal{V}_1(\omega,\omega,\mathcal{U}(\omega, C),C)\\
    \qquad+\mathcal{V}_1(\omega,\omega,\mathcal{U}_1(\omega,C,C))-\mathcal{V}_1(\omega,\mathcal{V}_1(\omega,\omega,C),C)=0\,,
\end{align*}
which can be rewritten in terms of symbols as 
\begin{align*}
    \begin{aligned}
    &-\exp(p_{01})V_1(p_0,p_2,p_3,p_4,p_5)+V_1(p_0,p_1,p_2,p_6)\mathcal{U}_1(y_6,p_3,p_4,p_5)-V_1(p_0,p_1,p_2+p_3,p_4,p_5)\\
    &+\exp(p_{34})V_1(p_0,p_1,p_2,p_4,p_5)-V_1(p_0,p_1,p_6,p_5)V_1(y_6,p_2,p_3,p_4)+V_1(p_0,p_1+p_2,p_3,p_4,p_5)=0\,.
    \end{aligned}
\end{align*}
Nesting one vertex into another is easy to evaluate thanks to the exponential form of the vertices. We find
\begin{align*}
    \begin{aligned}
    &V_1(p_0,p_1,p_2,p_6)\mathcal{U}_1(y_6,p_3,p_4,p_5)=\int_{0}^{1}dt_2\int_{0}^{t_2}dt_1\int_{0}^{1}ds_2\int_{0}^{s_2}ds_1p_{12}(t_1p_{13}+t_2p_{23})\\
    &\times\exp((1-t_1)p_{01}+(1-t_2)p_{02}+t_1(1-s_2)p_{14}+t_1s_2p_{15}+t_2(1-s_2)p_{24}+t_2s_2p_{25}+(1-s_1)p_{34}+s_1p_{35}),\\
    &V_1(p_0,p_1,p_6,p_5)V_1(y_6,p_2,p_3,p_4)=\int_{0}^{1}dt_2\int_{0}^{t_2}dt_1\int_{0}^{1}ds_2\int_{0}^{s_2}ds_1p_{23}((1-s_1)p_{12}+(1-s_2)p_{13})\\
    &\times\exp((1-t_1)p_{01}+(1-t_2)(1-s_1)p_{02}+(1-t_2)(1-s_2)p_{03}+t_1p_{15}+t_2(1-s_1)p_{25}+t_2(1-s_2)p_{35}\\
    &+s_1p_{24}+s_2p_{34}).
    \end{aligned}
\end{align*}
It is now easy to see that the consistency condition is satisfied order by order with $\mathcal{V}_1(\omega,\omega,C,C)=G_1(\omega,\omega,C,C)$. In particular, the integrals, after Taylor expansion, can easily be done and lead to simple rational numbers. We have checked enough consistency relations to support the expressions for the vertices.

\section{Quintic vertex}
\label{app:quintic}
Out of curiosity and to show the effectiveness and locality of the higher order vertices, let us compute the contribution to the quintic vertex $\mathcal{V}(\omega,\omega,C,C,C)$ that corresponds to the following graph
$$
   G= 
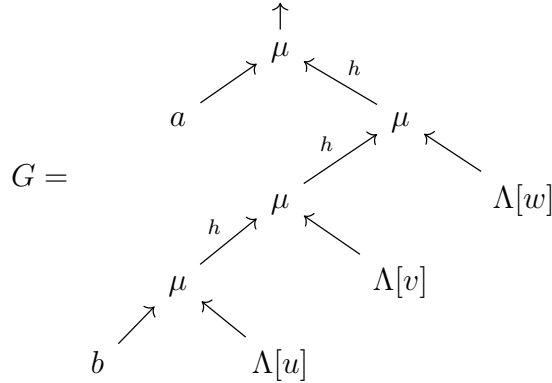
\begin{tikzcd}[column sep=small,row sep=small]
       && {} & & &\\
    && \mu\arrow[u] & & &\\
    &  a\arrow[ur]& &\mu\arrow[ul, "h" ']  &&\\
    &&  \mu\arrow[ur,"h"] &  &\arrow[ul]\Lambda[w]&&\\
    & \mu\arrow[ur,"h"] &&\arrow[ul]\Lambda[v]&&& \\
    b\arrow[ur] & & \arrow[ul]\Lambda[u] & &&&
\end{tikzcd}
$$
The final answer is a six-fold integral over the `times' $t_{i}$, $i=1,\ldots,6$:
\begin{align*}
    G&= (p_{12})^3\exp[(1-u_1-u_2-u_3) p_{01}+(1-u_4-u_5-u_6) p_{02}+u_1 p_{13}+u_2 p_{14}+\\
    &\qquad\qquad\qquad\qquad +u_3 p_{15}+u_4 p_{23}+u_5 p_{24}+u_6 p_{25}]\,,
\end{align*}
where the integration variables are expressed as
\begin{align*}
    u_1&=\frac{t_1 \left(t_2-1\right) t_3 t_4 \left(t_5-1\right) t_6}{-t_2 t_4 t_5+t_1 t_3 \left(\left(2 t_2-1\right) t_4 t_5-t_2\right)+1} \,, &
    u_2&=\frac{t_2 \left(t_1 t_3-1\right) t_4 \left(t_5-1\right) t_6}{-t_2 t_4 t_5+t_1 t_3 \left(\left(2 t_2-1\right) t_4 t_5-t_2\right)+1} \,,\\
    u_3&= \frac{\left(-t_2 t_4+t_1 t_3 \left(t_2 \left(2 t_4-1\right)-t_4\right)+1\right) t_5 t_6}{-t_2 t_4 t_5+t_1 t_3 \left(\left(2 t_2-1\right) t_4 t_5-t_2\right)+1}\,, &
    u_4&=\frac{t_1 \left(-t_2 t_3+\left(t_2 \left(2 t_3-1\right)-t_3\right) t_4 t_5+1\right)}{-t_2 t_4 t_5+t_1 t_3 \left(\left(2 t_2-1\right) t_4 t_5-t_2\right)+1} \,,\\
    u_5&= \frac{\left(t_1-1\right) t_2 \left(t_4 t_5-1\right)}{-t_2 t_4 t_5+t_1 t_3 \left(\left(2 t_2-1\right) t_4 t_5-t_2\right)+1}\,,&
    u_6&= \frac{\left(t_1-1\right) \left(t_2-1\right) t_5}{-t_2 t_4 t_5+t_1 t_3 \left(\left(2 t_2-1\right) t_4 t_5-t_2\right)+1}\,.
\end{align*}
As is clear, the vertex is local! It is also clear what the simplest form of the vertex at any order has to be. After a similar change of variables in the final expressions we get for $G_1$ and $G_2$ much more concise expressions
\begin{align*}
    G_1&=+(p_{12})^2 \exp\big[(1-u_1-u_2) p_{01}+(1-u_3-u_4) p_{02}+u_1 p_{13}+u_2 p_{14}+u_3 p_{23}+u_4 p_{24}\big]\,,\\
    G_2&= -(p_{13})^2\exp\big[(1-u_1-u_2) p_{01}+(1+u_3-u_4) p_{03}+u_1 p_{12}+u_2 p_{14}+u_3 p_{23}+u_4 p_{34}\big]\,.
\end{align*}
In particular, the Jacobian of the transformation eliminates the prefactors in $G_{1,2}$ of Appendix \ref{app:nnlo}. This shows that $G_1$ is related to $G_2$ via a simple permutation $\sigma_{23}$ followed by $u_3\rightarrow -u_3$. 

\footnotesize
\providecommand{\href}[2]{#2}\begingroup\raggedright\endgroup

\end{document}